\documentclass[aps,twocolumn,prb,floatfix,showpacs,superscriptaddress]{revtex4}

\usepackage{amsmath}
\usepackage{amssymb}
\usepackage{amsfonts}
\usepackage[pdftex]{graphicx}
\usepackage{units}
\usepackage{bm}
\usepackage{color}
\usepackage{tikz}
\usepackage{color}
\usepackage{array}

\definecolor{dunkelgrau}{rgb}{0.8,0.8,0.8}
\definecolor{hellgrau}{rgb}{0.95,0.95,0.95}
\newcommand{\cone}{\mathrm{i}}

\newcommand{\operator}[1]{\ensuremath{\hat{#1}}}
\newcommand{\mat}[1]{\boldsymbol{\mathsf{#1}}}
\newcommand{\euler}{\mathrm{e}}
\renewcommand{\vec}[1]{\boldsymbol{#1}}
\newcommand{\bra}[1]{\langle {#1} |}
\newcommand{\ket}[1]{| {#1} \rangle}

\begin{document}
\title{Non-local Gilbert damping tensor within the torque-torque correlation model}

\author{Danny Thonig}
\email{danny.thonig@physics.uu.se}
\affiliation{Department of Physics and Astronomy, Material Theory, Uppsala University, SE-75120 Uppsala, Sweden}

\author{Yaroslav Kvashnin}
\affiliation{Department of Physics and Astronomy, Material Theory, Uppsala University, SE-75120 Uppsala, Sweden}

\author{Olle Eriksson}
\affiliation{Department of Physics and Astronomy, Material Theory, Uppsala University, SE-75120 Uppsala, Sweden}
\affiliation{School of Science and Technology, \"Orebro University, SE-701 82 \"Orebro, Sweden}

\author{Manuel Pereiro}
\affiliation{Department of Physics and Astronomy, Material Theory, Uppsala University, SE-75120 Uppsala, Sweden}


\date{\today}

\begin{abstract}
An essential property of magnetic devices is the relaxation rate in magnetic switching which depends strongly on the damping in the magnetisation dynamics. It was recently measured that damping depends on the magnetic texture and, consequently, is a non-local quantity. The damping  enters the Landau-Lifshitz-Gilbert equation as the phenomenological Gilbert damping parameter $\alpha$, that does not, in a straight forward formulation, account for non-locality. Efforts were spent recently to obtain Gilbert damping from first principles for magnons of wave vector $\vec{q}$. However, to the best of our knowledge, there is no report about real space non-local Gilbert damping $\alpha_{ij}$. 
Here, a torque-torque correlation model based on a tight binding approach is applied to the bulk elemental itinerant magnets and it predicts significant off-site Gilbert damping contributions, that could be also negative.  Supported by atomistic magnetisation dynamics simulations we reveal the importance of the non-local Gilbert damping in atomistic magnetisation dynamics. This study gives a deeper understanding of the dynamics of the magnetic moments and dissipation processes in real magnetic materials. Ways of manipulating non-local damping are explored, either by temperature, material´s doping or strain. 
\end{abstract}









\maketitle

Efficient spintronics applications call for magnetic materials with low energy dissipation when moving magnetic textures, e.g. in race track memories \cite{Anonymous:mXuz0sqX}, skyrmion logics \cite{Iwasaki:2013hb,Fert:2013fq}, spin logics \cite{BehinAein:2010hj}, spin-torque nano-oscillator for neural network applications \cite{Anonymous:jd} or, more recently, soliton devices \cite{Koumpouras:2017uc}. In particular, the dynamics of such magnetic textures --- magnetic domain walls, magnetic Skyrmions, or magnetic solitons --- is well described in terms of precession and damping of the magnetic moment $\vec{m}_i$ as it is formulated in the atomistic Landau-Lifshitz-Gilbert (LLG) equation for site $i$

\begin{align}
	\frac{\partial \vec{m}_i}{\partial t}=\vec{m}_i\times\left(-\gamma \vec{B}_i^{eff}+\frac{\mat{\alpha}}{m_s}\frac{\partial \vec{m}_i}{\partial t}\right),
\end{align}
where $\gamma$ and $m_s$ are the gyromagnetic ratio and the magnetic moment length, respectively. The precession field $\vec{B}_i^{eff}$ is of quantum mechanical origin and is obtained either from effective spin-Hamilton models \cite{Eriksson:2016uw} or from first-principles \cite{Antropov:1996td}. In turn, energy dissipation is dominated by the ad-hoc motivated viscous damping in the equation of motion scaled by the Gilbert damping tensor $\mat{\alpha}$. Commonly, the Gilbert damping is used as a scalar parameter in magnetization dynamics simulations based on the LLG equation. Strong efforts were spend in the last decade to put the Gilbert damping to a first-principles ground derived for collinear magnetization configurations. Different methods were proposed: e.g. the breathing Fermi surface \cite{Kambersky:2007bp,Kambersky:1984iz,Kambersky:1976gi}, the torque-torque correlation \cite{Gilmore:2007eva}, spin-pumping \cite{Starikov:2010cz} or a linear response model \cite{Ebert:2011gx,Mankovsky:2013ii}. Within a certain accuracy, the theoretical models allow to interpret \cite{Schoen:2016gcc} and reproduce experimental trends \cite{Chico:2016dy, Durrenfeld:2015vx,Schoen:2017kv,Schoen:2017hj}.

Depending on the model, deep insight into the fundamental electronic-structure mechanism of the Gilbert damping $\alpha$ is provided: Damping is a Fermi-surface effect and depending on e.g. scattering rate, damping occurs due to spin-flip but also spin-conservative transition within a degenerated (intraband, but also interband transitions) and between non-degenerated (interband transitions) electron bands. As a consequence of these considerations, the Gilbert damping is proportional to the density of states, but it also scales with spin-orbit coupling\cite{Barati:2014gha,Pan:2016gb}. The scattering rate $\Gamma$ for the spin-flip transitions is allocated to thermal, but also correlation effects, making the Gilbert damping strongly temperature dependent which must be a consideration when applying a three-temperature model for the thermal baths, say phonon \cite{Ebert:2011gx}, electron, and spin temperature \cite{Thonig:2014kt}. In particular, damping is often related to the dynamics of a collective precession mode (macrospin approach) driven from an external perturbation field, as it is used in ferromagnetic resonance experiments (FMR) \cite{Ma:2017bd}. It is also established that the Gilbert damping depends on the orientation of the macrospin \cite{Steiauf:2005bv} and is, in addition, frequency dependent \cite{Thonig:2015ur}.

\begin{figure}
	\centering
	\includegraphics[width=0.75\linewidth]{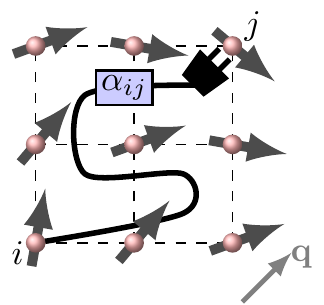}
	\caption{Schematic illustration of non-local energy dissipation $\alpha_{ij}$ between site $i$ and $j$ (red balls) represented by a power cord in a system with spin wave (gray arrows) propagation $\vec{q}$. }
	\label{fig:0}
\end{figure}

More recently, the role of non-collective modes to the Gilbert damping has been debated. F\"ahnle et al. \cite{Fahnle:2006cf} suggested to consider damping in a tensorial and non-isotropic form via $\mat{\alpha}_i$ that differs for different sites $i$ and depends on the whole magnetic configuration of the system. As a result, the experimentally and theoretically assumed local Gilbert equation is replaced by a non-local equation via non-local Gilbert damping $\alpha_{ij}$ accounting for the most general form of Rayleigh's dissipation function\cite{Minguzzi:2015jsd}. The proof of principles was given for magnetic domain walls \cite{Yuan:2014en,Nembach:2013wh}, linking explicitly the Gilbert damping to the gradients in the magnetic spin texture $\nabla \vec{m}$. Such spatial non-locality, in particular, for discrete atomistic models, allows further to motivate energy dissipation between two magnetic moments at sites $i$ and $j$, and is represented by $\alpha_{ij}$, as schematically illustrated in Fig. \ref{fig:0}. An analytical expression for $\alpha_{ij}$ was already proposed by various authors \cite{Ebert:2011gx,Bhattacharjee:2012if,Gilmore:2009hm}, however, not much work has been done on a material specific, first-principle description of the atomistic non-local Gilbert damping $\alpha_{ij}$. An exception is the work by Gilmore et al. \cite{Gilmore:2009hm} who studied $\alpha(\vec{q})$ in the reciprocal space as a function of the magnon wave vector $\vec{q}$ and concluded that the non-local damping is negligible.
Yan et al. \cite{Yuan:2014en} and Hals et al.\cite{Hals:2009gy}, on the other hand, applied scattering theory according to Brataas et al.\cite{Brataas:2011caa} to simulate non-collinearity in Gilbert damping, only in reciprocal space or continuous mesoscopic scale.  Here we come up with a technical description of non-locality of the damping parameter $\alpha_{ij}$, in real space, and provide numerical examples for elemental, itinerant magnets, which might be of high importance in the context of ultrafast demagnetization \cite{Chimata:2017iu}.

The paper is organized as follows: In Section~\ref{sec:method}, we introduce our first-principles model formalism based on the torque-torque correlation model to study non-local damping. This is applied to bulk itinerant magnets bcc Fe, fcc Co, and fcc Ni in both reciprocal and real space and it is analysed in details in Section~\ref{sec:relrecip}. Here, we will also apply atomistic magnetisation dynamics to outline the importance in the evolution of magnetic systems. Finally, in the last section, we conclude the paper by giving an outlook of our work.

\section{Methods}
\label{sec:method}

We consider the torque-torque correlation model introduced by Kambersk\'y \cite{Kambersky:1984iz} and further elaborated on by Gilmore et al. \cite{Gilmore:2007eva}. Here, finite magnetic moment rotations couple to the Bloch eigenenergies $\varepsilon_{n,\vec{k}}$ and eigenstates $\ket{n\vec{k}}$, characterised by the band index $n$ at wave vector $\vec{k}$, due to spin-orbit coupling. This generates a non-equilibrium population state (a particle-hole pair), where the excited states relax towards the equilibrium distribution (Fermi-Dirac statistics) within the time $\tau_{n,\vec{k}}=\nicefrac{1}{\Gamma}$, which we assume is independent of $n$ and $\vec{k}$. In the adiabatic limit, this perturbation is described by the Kubo-Greenwood perturbation theory and reads \cite{Gilmore:2007eva,Gilmore2008:tw} in a non-local formulation

\begin{align}
	\alpha^{\mu\nu}(\vec{q})=\frac{g\pi}{m_s}\int_\Omega \sum_{nm} T^\mu_{n\vec{k};m\vec{k}+\vec{q}} \left(T^\nu_{n\vec{k};m\vec{k}+\vec{q}}\right)^* W_{n\vec{k};m\vec{k}+\vec{q}} \mathrm{d}\vec{k}.
	\label{eq:1.1}
\end{align}

Here the integral runs over the whole Brillouin zone volume $\Omega$. A frozen magnon of wave vector $\vec{q}$ is considered that is ascribed to the non-locality of $\alpha$. The scattering events depend on the spectral overlap $W_{n\vec{k};m\vec{k}+\vec{q}}=\int \eta(\varepsilon) A_{n\vec{k}}\left(\varepsilon,\Gamma\right)A_{m\vec{k}+\vec{q}}\left(\varepsilon,\Gamma\right)\mathrm{d}\varepsilon$ between two bands $\varepsilon_{n,\vec{k}}$ and $\varepsilon_{m,\vec{k}+\vec{q}}$, where the spectral width of the electronic bands $A_{n\vec{k}}$ is approximated by a Lorentzian of width $\Gamma$. Note that $\Gamma$ is a parameter in our model and can be spin-dependent as proposed in Ref.~[\onlinecite{Gilmore:2011be}]. In other studies, this parameter is allocated to the self-energy of the system and is obtained by introducing disorder, e.g., in an alloy or alloy analogy model using the coherent potential approximation\cite{Ebert:2011gx} (CPA) or via the inclusion of electron correlation \cite{Sayad:2016dw}. Thus, a principle study of the non-local damping versus $\Gamma$ can be also seen as e.g. a temperature dependent study of the non-local damping. $\eta=\nicefrac{\partial f}{\partial \varepsilon}$ is the derivative of the Fermi-Dirac distribution $f$ with respect to the energy. $T^\mu_{n\vec{k},m\vec{k}+\vec{q}}=\bra{n\vec{k}}\operator{T}^\mu\ket{m\vec{k}+\vec{q}}$, where $\mu=x,y,z$, are the matrix elements of the torque operator $\vec{\operator{T}}=\left[\vec{\sigma},\mathcal{H}_{so}\right]$ obtained from variation of the magnetic moment around certain rotation axis $\vec{e}$. $\vec{\sigma}$ and $\mathcal{H}_{so}$ are the Pauli matrices and the spin-orbit hamiltonian, respectively. In the collinear ferromagnetic limit, $\vec{e}=\vec{e}_z$ and variations occur in $x$ and $y$, only, which allows to consider just one component of the torque, i.e. $\operator{T}^-=\operator{T}^x-\cone \operator{T}^y$. Using Lehmann representation \cite{Zabloudil:2005dJ}, we rewrite the Bloch eigenstates by Green's function $\mathcal{G}$, and define the spectral function $\operator{A}=\cone\left(\mathcal{G}^R-\mathcal{G}^A\right)$ with the retarded (R) and advanced (A) Green's function,

\begin{align}
	\alpha^{\mu\nu}(\vec{q})=\frac{g}{m\pi}\int \int_\Omega \eta(\varepsilon) \operator{T}^\mu \operator{A}_{\vec{k}}\left(\operator{T}^\nu\right)^\dagger \operator{A}_{\vec{k}+\vec{q}} \mathrm{d}\vec{k}\mathrm{d}\varepsilon.
		\label{eq:1.2}
\end{align}

The Fourier transformation of the Green's function $\mathcal{G}$ finally is used to obtain the non-local Gilbert damping tensor \cite{Thonig:2014kt} between site $i$ at position $\vec{r}_i$ and site $j$ at position $\vec{r}_j$,

\begin{align}
	\alpha_{ij}^{\mu\nu}=\frac{g}{m\pi} \int \eta(\varepsilon) \operator{T}_i^\mu\operator{A}_{ij}\left(\operator{T}_j^\nu\right)^\dagger \operator{A}_{ji}\mathrm{d}\varepsilon.
		\label{eq:1.3}
\end{align}

Note that $\operator{A}_{ij}=\cone\left(\mathcal{G}^R_{ij}-\mathcal{G}^A_{ji}\right)$. This result is consistent with the formulation given in Ref.~[\onlinecite{Bhattacharjee:2012if}] and Ref.~[\onlinecite{Ebert:2011gx}]. Hence, the definition of non-local damping in real space and reciprocal space translate into each other by a Fourier transformation, 

\begin{align}
	\alpha_{ij}=\int \alpha\left(\vec{q}\right)\euler^{-\cone (\vec{r}_j-\vec{r}_i) \cdot \vec{q}}\mathrm{d}\vec{q}.
	\label{eq:1.4}
\end{align}

Note the obvious advantage of using Eq.~\eqref{eq:1.3}, since it allows for a direct calculation of $\alpha_{ij}$, as opposed to taking the inverse Fourier transform of Eq.~\eqref{eq:1.4}. For first-principles studies, the Green's function is obtained from a tight binding (TB) model based on the Slater-Koster parameterization \cite{Slater:1954hi}. The Hamiltonian consists of on-site potentials, hopping terms, Zeeman energy, and spin-orbit coupling (See Appendix \ref{sec:AppendixA}). The TB parameters, including the spin-orbit coupling strength, are obtained by fitting the TB band structures to ab initio band structures as reported elsewhere \cite{Thonig:2014kt}. 

\begin{figure*}
	\centering
	\includegraphics[width=0.9\textwidth]{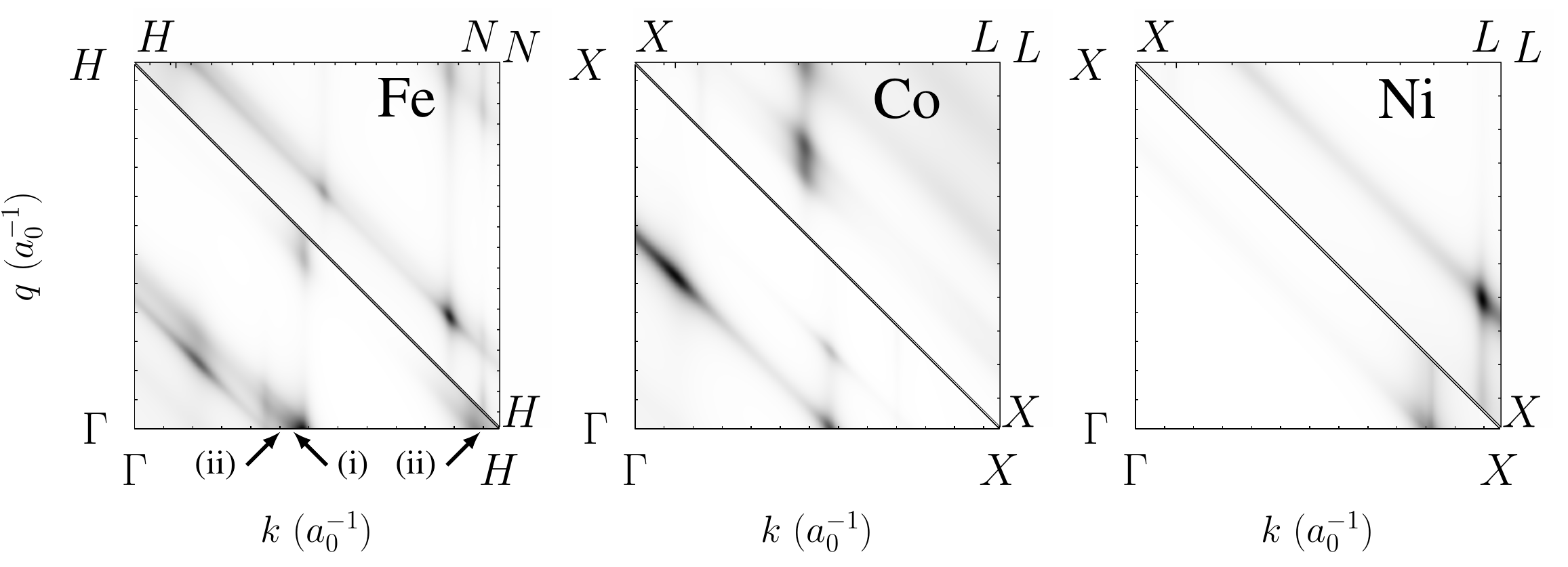}
	\caption{Electronic state resolved non-local Gilbert damping obtained from the integrand of Eq.~\eqref{eq:1.2} along selected paths in the Brillouin zone for bcc Fe, fcc Co and fcc Ni. The scattering rate used is $\Gamma=\unit[0.01]{eV}$. The abscissa (both top and bottom in each panels) shows the momentum path of the electron $\vec{k}$, where the ordinate (left and right in each panel) shows the magnon propagation vector $\vec{q}$. The two `triangle' in each panel should be viewed separately where the magnon momentum changes accordingly (along the same path) to the electron momentum. }
	\label{fig:1}
\end{figure*}

Beyond our model study, we simulate material specific non-local damping with the help of the full-potential linear muffin-tin orbitals (FP-LMTO) code ``RSPt" \cite{Wills:1987kc,Dreysse:2000uv}. Further numerical details are provided in Appendix \ref{sec:AppendixA}.

With the aim to emphasize the importance of non-local Gilbert damping in the evolution of atomistic magnetic moments, we performed atomistic magnetization dynamics by numerical solving the Landau-Lifshitz Gilbert (LLG) equation, explicitly incorporating non-local damping \cite{Brataas:2011caa,Vittoria:2010ft,Thonig:2014kt}

\begin{align}
	\frac{\partial \vec{m}_i}{\partial t}=\vec{m}_i\times\left(-\gamma \vec{B}_i^{eff}+\sum_j\frac{\mat{\alpha}_{ij}}{m^j_s}\frac{\partial \vec{m}_j}{\partial t}\right).
    \label{eq:1.5}
\end{align}

\noindent Here, the effective field $\vec{B}_i^{eff}=-\nicefrac{\partial \operator{H}}{\partial \vec{m}_i}$ is allocated to the spin Hamiltonian entails Heisenberg-like exchange coupling $-\sum_{ij}J_{ij}\vec{m}_i\cdot\vec{m}_j$ and uniaxial magneto-crystalline anisotropy $\sum_i K_i(\vec{m}_i\cdot\vec{e}_i)^2$ with the easy axis along $\vec{e}_i$. $J_{ij}$ and $K_i$ are the Heisenberg exchange coupling and the magneto-crystalline anisotropy constant, respectively, and were obtained from first principles \cite{Bottcher:2012hz,Szilva:2013fo}. Further details are provided in Appendix~\ref{sec:AppendixA}.

\section{Results and Discussion}
\label{sec:relrecip}

This section is divided in three parts. In the first part, we discuss non-local damping in reciprocal space $\vec{q}$. The second part deals with the real space definition of the Gilbert damping $\alpha_{ij}$. Atomistic magnetization dynamics including non-local Gilbert damping is studied in the third part.

\subsection{Non-local damping in reciprocal space}
\label{susec:recip}

The formalism derived by Kambersk\'y \cite{Kambersky:1984iz} and Gilmore \cite{Gilmore:2007eva} in Eq.~\eqref{eq:1.1} represents the non-local contributions to the energy dissipation in the LLG equation by the magnon wave vector $\vec{q}$. In particular, Gilmore et al. \cite{Gilmore:2009hm} concluded that for transition metals at room temperature the single-mode damping rate is essentially independent of the magnon wave vector for $\vec{q}$ between $0$ and $1\%$ of the Brillouin zone edge. However, for very small scattering rates $\Gamma$, Gilmore and Stiles \cite{Gilmore:2007eva} observed for bcc Fe, hcp Co and fcc Ni a strong decay of $\alpha$ with $\vec{q}$, caused by the weighting function $W_{nm}(\vec{k},\vec{k}+\vec{q})$ without any significant changes of the torque matrix elements. Within our model systems, we observed the same trend for bcc Fe, fcc Co and fcc Ni. To understand the decay of the Gilbert damping with magnon-wave vector $\vec{q}$ in more detail, we study selected paths of both the magnon $\vec{q}$ and electron momentum $\vec{k}$ in the Brillouin zone at the Fermi energy $\varepsilon_F$ for bcc Fe ($\vec{q},\vec{k}\in\Gamma\rightarrow H$ and $\vec{q},\vec{k}\in H\rightarrow N$), fcc Co and fcc Ni ($\vec{q},\vec{k}\in\Gamma\rightarrow X$ and $\vec{q},\vec{k}\in X\rightarrow L$) (see Fig.~\ref{fig:1}, where the integrand of Eq.~\eqref{eq:1.1} is plotted). For example, in Fe, a usually two-fold degenerated $d$ band (approximately in the middle of $\overline{\Gamma H}$, marked by $(i)$) gives a significant contribution to the intraband damping for small scattering rates.  There are two other contributions to the damping (marked by $(ii)$), that are caused purely by interband transitions. With increasing, but small $\vec{q}$ the intensities of the peaks decrease and interband transitions become more likely. With larger $\vec{q}$, however, more and more interband transitions appear which leads to an increase of the peak intensity, significantly in the peaks marked with $(ii)$. This increase could be the same order of magnitude as the pure intraband transition peak. Similar trends also occur in Co as well as Ni and are also observed for Fe along the path $\overline{HN}$. Larger spectral width $\Gamma$ increases the interband spin-flip transitions even further (data not shown).  Note that the torque-torque correlation model might fail for large values of $\vec{q}$, since the magnetic moments change so rapidly in space that the adiababtic limit is violated\cite{Garate:2009ija} and electrons are not stationary equilibrated. The electrons do not align according the magnetic moment and the non-equilibrium electron distribution in Eq.~\eqref{eq:1.1} will not fully relax. In particular, the magnetic force theorem used to derive Eq.~\eqref{eq:1.2} may not be valid.  

\begin{figure}
	\centering
	\includegraphics[width=1.05\linewidth]{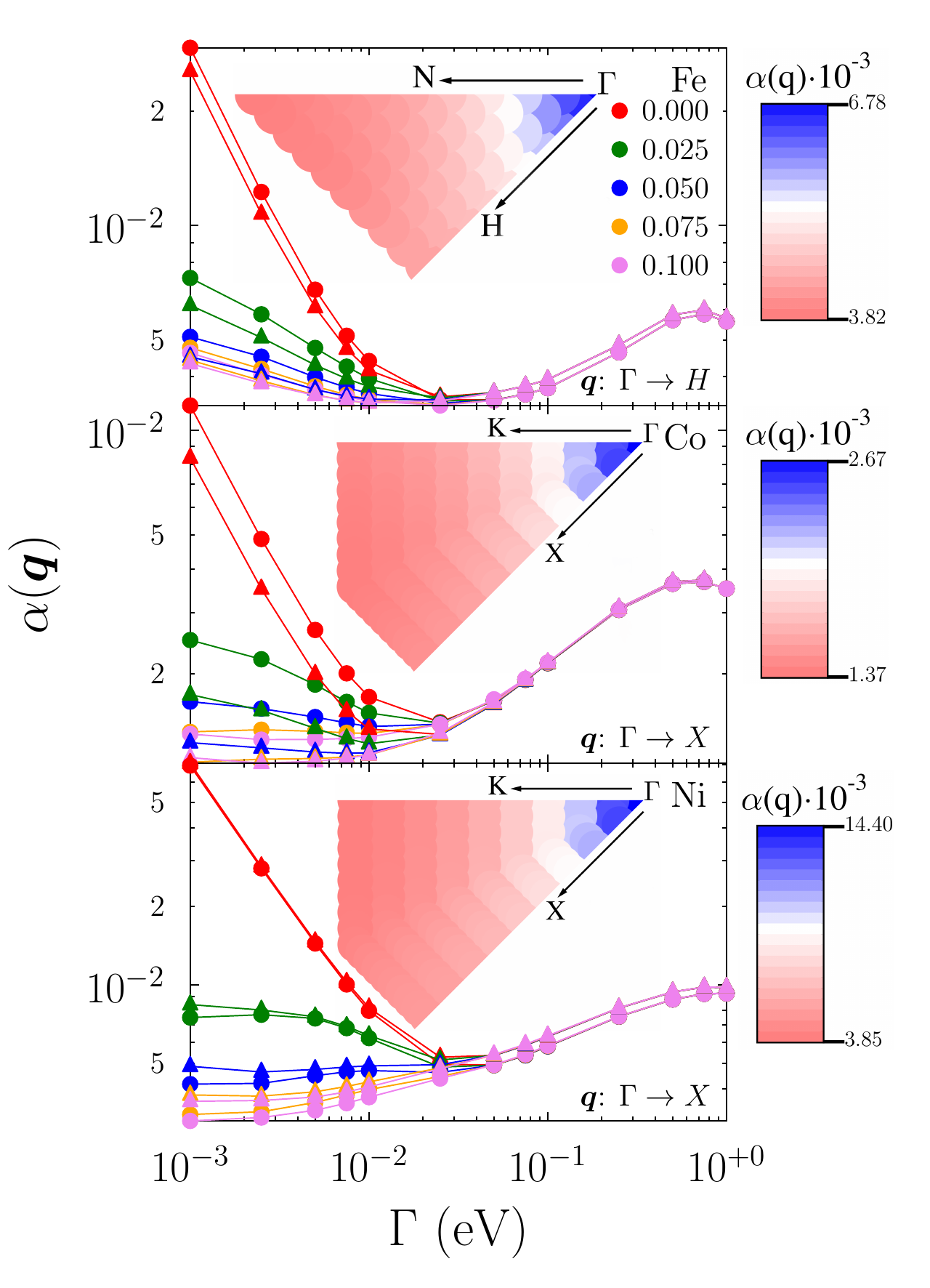}
	\caption{(Color online) Non-local Gilbert damping as a function of the spectral width $\Gamma$ for different reciprocal wave vector $q$ (indicated by different colors and in units $a_0^{-1}$). Note that $q$ provided here are in direct coordinates and only the direction differs between the different elementals, itinerant magnets. The non-local damping is shown for bcc Fe (top panel) along $\Gamma\rightarrow H$, for fcc Co (middle panel) along $\Gamma\rightarrow X$, and for fcc Ni (bottom panel) along $\Gamma\rightarrow X$.  It is obtained from `Lorentzian' (Eq.~\eqref{eq:1.1}, circles) and Green's function (Eq.~\eqref{eq:1.2}, triangles) method. The directional dependence of $\alpha$ for $\Gamma=\unit[0.01]{eV}$ is shown in the inset.}
	\label{fig:2}
\end{figure}

The integration of the contributions in electron momentum space $\vec{k}$ over the whole Brillouin zone is presented in Fig.~\ref{fig:2}, where both `Loretzian' method given by Eq.~\eqref{eq:1.1} and Green's function method represented by Eq.~\eqref{eq:1.2} are applied. Both methods give the same trend, however, differ slightly in the intraband region, which was already observed previously by the authors of Ref.~[\onlinecite{Thonig:2014kt}]. In the `Lorentzian' approach, Eq.~\eqref{eq:1.1}, the electronic structure itself is unaffected by the scattering rate $\Gamma$, only the width of the Lorentian used to approximate $A_{n\vec{k}}$ is affected. In the Green function approach, however, $\Gamma$ enters as the imaginary part of the energy at which the Green functions is evaluated and, consequently, broadens and shifts maxima in the spectral function. This offset from the real energy axis provides a more accurate description with respect to the ab initio results than the Lorentzian approach.

Within the limits of our simplified electronic structure tight binding method, we obtained qualitatively similar trends as observed by Gilmore et al. \cite{Gilmore:2009hm}: a dramatic decrease in the damping at low scattering rates $\Gamma$ (intraband region). This trend is common for all here observed itinerant magnets typically in a narrow region $0<\left|\vec{q}\right|<0.02a^{-1}_0$, but also for different magnon propagation directions. For larger $\left|\vec{q}\right|>0.02a^{-1}_0$ the damping could again increase (not shown here). The decay of $\alpha$ is only observable below a certain threshold scattering rate $\Gamma$, typically where intra- and interband contribution equally contributing to the Gilbert damping. As already found by Gilmore et al. \cite{Gilmore:2009hm} and Thonig et al.\cite{Thonig:2014kt}, this point is materials specific. In the interband regime, however, damping is independent of the magnon propagator, caused by already allowed transition between the electron bands due to band broadening. Marginal variations in the decay with respect to the direction of $\vec{q}$ (Inset of Fig.~\ref{fig:2}) are revealed, which was not reported before. Such behaviour is caused by the break of the space group symmetry due to spin-orbit coupling and a selected global spin-quantization axis along z-direction, but also due to the non-cubic symmetry of $\mathcal{G}_{\vec{k}}$ for $\vec{k}\neq 0$.  As a result, e.g., in Ni the non-local damping decays faster along $\overline{\Gamma K}$ than in $\overline{\Gamma X}$. This will be discussed more in detail in the next section.

We also investigated the scaling of the non-local Gilbert damping with respect to the spin-orbit coupling strength $\xi_d$ of the d-states (see Appendix \ref{sec:AppendixB}). We observe an effect that previously has not been discussed, namely that the non-local damping has a different exponential scaling with respect to the spin-orbit coupling constant for different $\left|\vec{q}\right|$. In the case where $\vec{q}$ is close to the Brillouin zone center (in particular $\vec{q}=0$), $\alpha\propto \xi_d^3$ whereas for wave vectors $\left|\vec{q}\right|>0.02a^{-1}_0$, $\alpha\propto \xi_d^2$. For large $\vec{q}$, typically interband transitions dominate the scattering mechanism, as we show above and which is known to scale proportional to $\xi^2$. Here in particular, the $\xi^2$ will be caused only by the torque operator in Eq.~\eqref{eq:1.1}. On the other hand, this indicates that spin-mixing transitions become less important because there is not contribution in $\xi$ from the spectral function entering to the damping $\alpha(\vec{q})$. 

The validity of the Kambserk\'y model becomes arguable for $\xi^3$ scaling, as it was already proved by Costa et al. \cite{Costa:2015hp} and Edwards \cite{Edwards:2016be}, since it causes the unphysical and strong diverging intraband contribution at very low temperature (small $\Gamma$). Note that there is no experimental evidence of such a trend, most likely due to that sample impurities also influence $\Gamma$. Furthermore, various other methods postulate that the Gilbert damping for $\vec{q}=0$ scales like $\xi^2$ ~\cite{Mankovsky:2013ii,Kambersky:2007bp,Pan:2016gb}. Hence, the current applied theory, Eq.~\eqref{eq:1.2}, seems to be valid only in the long-wave limit, where we found $\xi^2$-scaling. On the other hand, Edwards \cite{Edwards:2016be} proved that the long-wave length limit ($\xi^2$-scaling) hold also in the short-range limit if one account only for transition that conserve the spin (`pure' spin states), as we show for Co in Fig.~\ref{fig:app:2} of Appendix \ref{sec:AppendixC}. The trends $\alpha$ versus $\left|\vec{q}\right|$  as described above changes drastically for the `corrected' Kambersk\'y formula: the interband region is not affected by these corrections. In the intraband region, however, the divergent behaviour of $\alpha$ disappears and the Gilbert damping monotonically increases with larger magnon wave vector and over the whole Brillouin zone. This trend is in good agreement with Ref.~[\onlinecite{Yuan:2014en}]. For the case, where $\vec{q}=0$, we even reproduced the results reported in Ref.~[\onlinecite{Barati:2014gha}]; in the limit of small scattering rates the damping is constant, which was also reported before in experiment \cite{Oogane:2006bz,Bhagat:1974iu}. Furthermore, the anisotropy of $\alpha(\vec{q})$ with respect to the direction of $\vec{q}$ (as discussed for the insets of Fig.~\ref{fig:2}) increases by accounting only for pure-spin states (not shown here). Both agreement with experiment and previous theory motivate to consider $\xi^2$-scaling for all $\Gamma$.

\subsection{Non-local damping in real space}
\label{susec:real}

\begin{figure}
	\centering
	\includegraphics[scale=0.8]{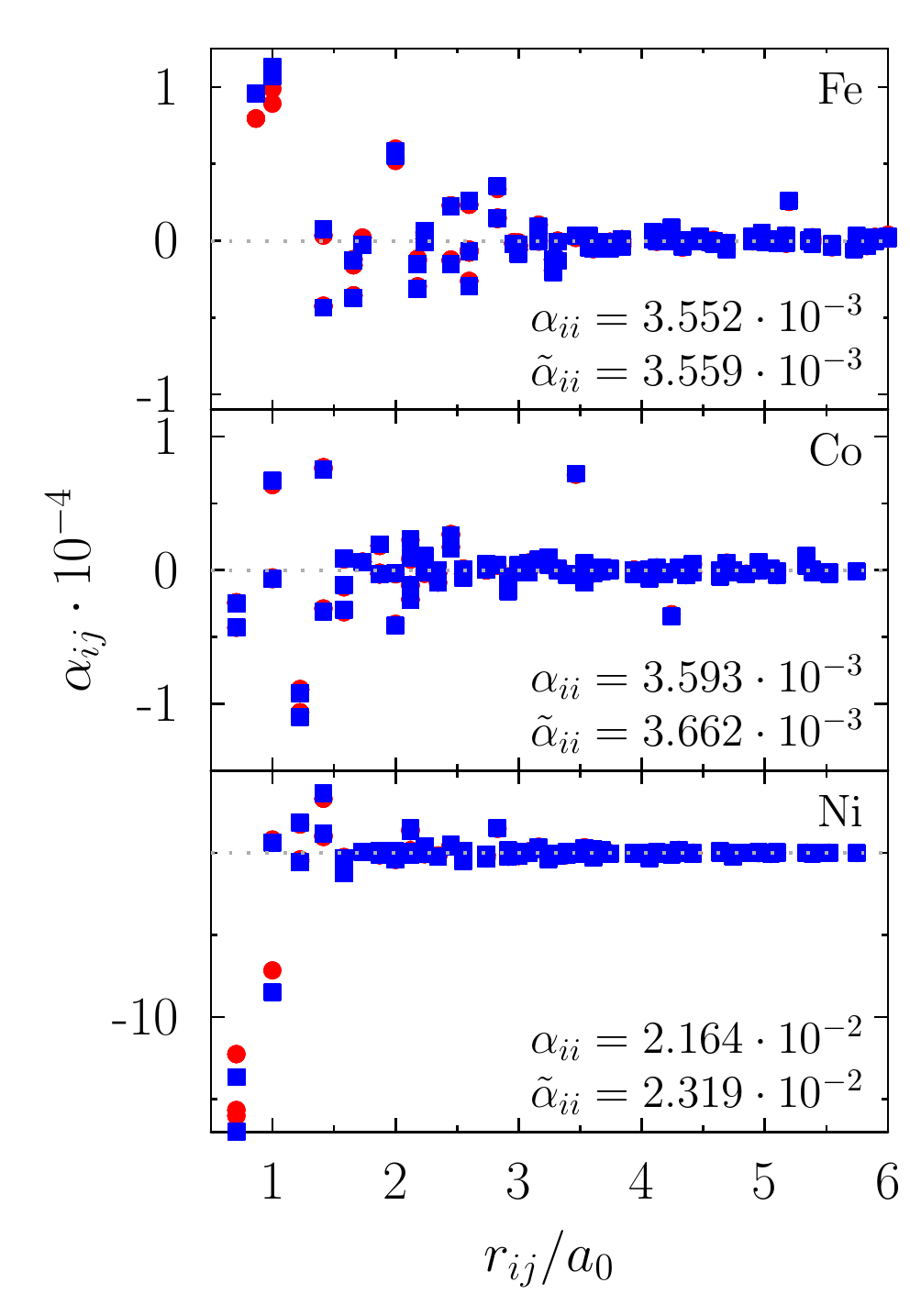}
	\caption{(Color online) Real-space Gilbert damping $\alpha_{ij}$ as a function of the distance $r_{ij}$ between two sites $i$ and $j$ for bcc Fe, fcc Co, and fcc Ni. Both the `corrected' Kambersk\'y (red circles) and the Kambersk\'y (blue squares) approach is considered. The distance is normalised to the lattice constant $a_0$. The on-site damping $\alpha_{ii}$ is shown in the figure label. The grey dotted line indicates the zero line. The spectral width is $\Gamma=\unit[0.005]{eV}$.}
	\label{fig:3}
\end{figure}

Atomistic spin-dynamics, as stated in Section~\ref{sec:method} (see Eq.~\eqref{eq:1.5}), that includes non-local damping requires Gilbert damping in real-space, e.g. in the form $\alpha_{ij}$. This point is addressed in this section. Such non-local contributions are not excluded in the Rayleigh dissipation functional, applied by Gilbert to derive the dissipation contribution in the equation of motion \cite{Gilbert:2004jg} (see Fig.~\ref{fig:3}).
                                                              
\begin{figure}
	\centering
	\includegraphics[scale=0.7]{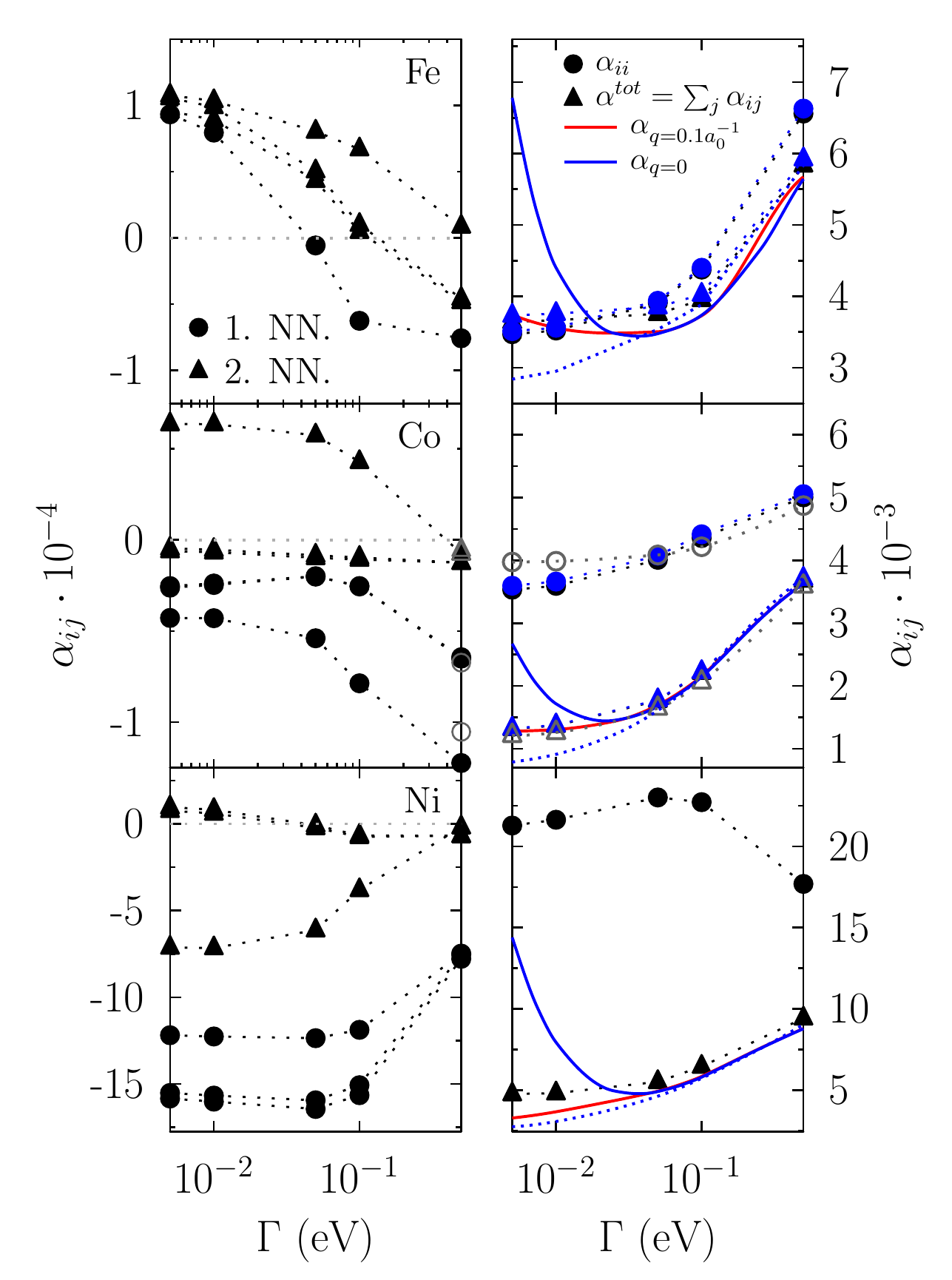}
	\caption{(Color online) First (circles) and second nearest neighbour (triangles) Gilbert damping (left panel) as well as on-site (circles) and total Gilbert (right panel) as a function of the spectral width $\Gamma$ for the itinerant magnets Fe, Co, and Ni. In particular for Co, the results obtained from tight binding are compared with first-principles density functional theory results (gray open circles). Solid lines (right panel) shows the Gilbert damping obtained for the magnon wave vectors $q=0$ (blue line) and $q=0.1 a_0^{-1}$ (red line). Dotted lines are added to guide the eye. Note that since cubic symmetry is broken (see text), there are two sets of nearest neighbor parameters and two sets of next nearest neighbor parameters (left panel) for any choice of $\Gamma$. }
	\label{fig:4}
\end{figure}

Dissipation is dominated by the on-site contribution $\alpha_{ii}$ in the itinerant magnets investigated here. For both Fe ($\alpha_{ii}=3.55\cdot 10^{-3}$) and Co ($\alpha_{ii}=3.59\cdot 10^{-3}$) the on-site damping contribution is similar, whereas for Ni $\alpha_{ii}$ is one order of magnitude higher. Off-site contributions $i\neq j$ are one-order of magnitude smaller than the on-site part and can be even negative. Such negative damping is discernible also in Ref.~[\onlinecite{Umetsu:2012ga}], however, it was not further addressed by the authors. Due to the presence of the spin-orbit coupling and a preferred global spin-quantization axis (in $z$-direction), the cubic symmetry of the considered itinerant magnets is broken and, thus, the Gilbert damping is anisotropic with respect to the sites $j$ (see also Fig.~\ref{fig:4} left panel). For example, in Co, four of the in-plane nearest neighbours (NN) are $\alpha_{NN}\approx -4.3\cdot 10^{-5}$, while the other eight are $\alpha_{NN}\approx -2.5\cdot 10^{-5}$. However, in Ni the trend is opposite: the out-of-plane damping ($\alpha_{NN}\approx -1.6\cdot 10^{-3}$) is smaller than the in-plane damping ($\alpha_{NN}\approx -1.2\cdot 10^{-3}$). Involving more neighbours, the magnitude of the non-local damping is found to decay as $\nicefrac{1}{r^2}$ and, consequently, it is different than the Heisenberg exchange parameter that asymptotically decays in RKKY-fashion as $J_{ij}\propto\nicefrac{1}{r^3}$~~\cite{Pajda:2001ix}. For the Heisenberg exchange, the two Green's functions as well as the energy integration in the Lichtenstein-Katsnelson-Antropov-Gubanov formula \cite{Liechtenstein:1987br} scales like $r_{ij}^{-1}$,
\begin{align}
	\mathcal{G}^\sigma_{ij}\propto\frac{e^{\cone(\vec{k}^\sigma\cdot\vec{r}_{ij}+\Phi^\sigma)}}{\left|\vec{r}_{ij}\right|}
\end{align}
 whereas for simplicity we consider here a single-band model but the results can be generalized also to the multiband case and where $\Phi^\sigma$ denotes a phase factor for spin $\sigma=\uparrow,\downarrow$. For the non-local damping the energy integration is omitted due to the properties of $\eta$ in Eq.~\eqref{eq:1.3} and, thus,
\begin{align}
	\alpha_{ij}\propto\frac{\sin\left[\vec{k}^\uparrow\cdot\vec{r}_{ij}+\Phi^\uparrow\right]\sin\left[\vec{k}^\downarrow\cdot\vec{r}_{ij}+\Phi^\downarrow\right]}{\left|\vec{r}_{ij}\right|^2}.
\end{align} 
This spatial dependency of $\alpha_{ij}$ superimposed with Ruderman-Kittel-Kasuya-Yosida (RKKY) oscillations was also found in Ref.~[\onlinecite{Umetsu:2012ga}] for a model system.

For Ni, dissipation is very much short range, whereas in Fe and Co `damping peaks' also occur at larger distances (e.g. for Fe at $r_{ij}=5.1 a_0$ and for Co at $r_{ij}=3.4 a_0$). The `long-rangeness' depends strongly on the parameter $\Gamma$ (not shown here). As it was already observed for the Heisenberg exchange interaction $J_{ij}$ \cite{Bottcher:2012hz}, stronger thermal effects represented by $\Gamma$ will reduce the correlation length between two magnetic moments at site $i$ and $j$. The same trend is observed for damping: larger $\Gamma$ causes smaller dissipation correlation length and, thus, a faster decay of non-local damping in space $r_{ij}$. Different from the Heisenberg exchange, the absolute value of the non-local damping typically decreases with $\Gamma$ as it is demonstrated in Fig.~\ref{fig:4}.

Note that the change of the magnetic moment length is not considered in the results discussed so far. The anisotropy with respect to the sites $i$ and $j$ of the non-local Gilbert damping continues in the whole range of the scattering rate $\Gamma$ and is controlled by it. For instance, the second nearest neighbours damping in Co and Ni become degenerated at $\Gamma=\unit[0.5]{eV}$, where the anisotropy between first-nearest neighbour sites increase. Our results show also that the sign of $\alpha_{ij}$ is affected by $\Gamma$ (as shown in Fig.~\ref{fig:4} left panel). Controlling the broadening of Bloch spectral functions $\Gamma$ is in principal possible to evaluate from theory, but more importantly it is accessible from experimental probes such as angular resolved photoelectron spectroscopy and two-photon electron spectroscopy. 

The importance of non-locality in the Gilbert damping depend strongly on the material (as shown in Fig.~\ref{fig:4} right panel). It is important to note that the total --- defined as $\alpha^{tot}=\sum_j \alpha_{ij}$ for arbitrary $i$ ---, but also the local ($i=j$) and the non-local ($i\neq j$) part of the Gilbert damping do not violate the thermodynamic principles by gaining angular momentum (negative total damping). For Fe, the local and total damping  are of the same order for all $\Gamma$, where in Co and Ni the local and non-local damping are equally important. The trends coming from our tight binding electron structure were also reproduced by our all-electron first-principles simulation, for both dependency on the spectral broadening $\Gamma$ (Fig.~\ref{fig:4} gray open circles) but also site resolved non-local damping in the intraband region (see Appendix \ref{sec:AppendixA}), in particular for fcc Co. 

\begin{figure}
	\centering
	\includegraphics[scale=0.8]{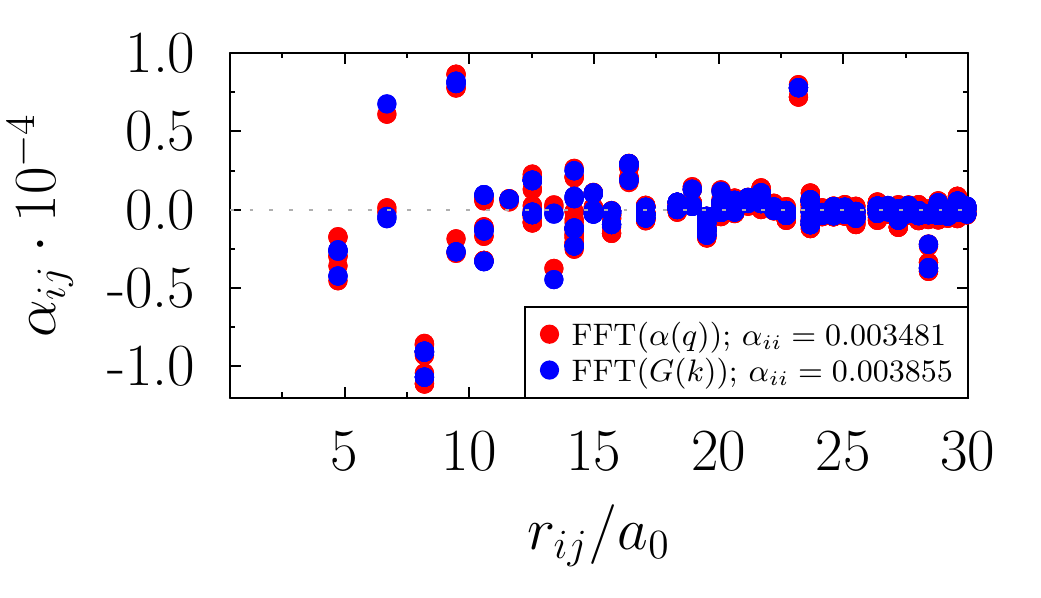}
	\caption{(Color online) Comparing non-local Gilbert damping obtained by Eq.~\eqref{eq:1.4} (red symbols) and Eq.~\eqref{eq:1.3} (blue symbols) in fcc Co for $\Gamma=\unit[0.005]{eV}$. The dotted line indicates zero value. }
	\label{fig:5}
\end{figure}

We compare also the non-local damping obtain from the real and reciprocal space. For this, we used Eq.~(\ref{eq:1.2}) by simulating $N_{\vec{q}}=15 \times 15 \times 15$ points in the first magnon Brillouin zone $\vec{q}$ and Fourier-transformed it (Fig.~\ref{fig:5}). For both approaches, we obtain good agreement, corroborating our methodology and possible applications in both spaces. The  non-local damping for the first three nearest neighbour shells turn out to converge rapidly with $N_{\vec{q}}$, while it does not converge so quickly for larger distances $\vec{r}_{ij}$. The critical region around the $\Gamma$-point in the Brillouin zone is suppressed in the integration over $\vec{q}$. On the other hand, the relation $\alpha^{tot}=\sum_j \alpha_{ij}=\alpha(q=0)$ for arbitrary $i$ should be valid, which is however violated in the intraband region as shown in Fig.~\ref{fig:4} (compare triangles and blue line in Fig.~\ref{fig:4}): The real space damping is constant for small $\Gamma$ and follows the long-wavelength limit (compare triangles and red line in Fig.~\ref{fig:4}) rather than the divergent ferromagnetic mode ($q=0$). Two explanations are possible: \textit{i)} convergence with respect to the real space summation and \textit{ii)} a different scaling in both models with respect to the spin-orbit coupling. For \textit{i)}, we carefully checked the convergence with the summation cut-off (see Appendix \ref{sec:AppendixD}) and found even a lowering of the total damping for larger cut-off. However, the non-local damping is very long-range and, consequently, convergence  will be achieved only at a cut-off radius $>>9a_0$. 

For \textit{ii)}, we checked the scaling of the real space Gilbert damping with the spin-orbit coupling of the $d$-states (see Appendix \ref{sec:AppendixB}). Opposite to the `non-corrected' Kambersk\'y formula in reciprocal space, which scales like $\xi_d^3$, we find $\xi_d^2$ for the real space damping. This indicates that the spin-flip scattering hosted in the real-space Green's function is suppressed. To corroborate this statement further, we applied the corrections proposed by Edwards \cite{Edwards:2016be} to our real space formula Eq.~\eqref{eq:1.3}, which by default assumes $\xi^2$ (Fig.~\ref{fig:3}, red dots). Both methods, corrected and non-corrected Eq.~\eqref{eq:1.3}, agree quite well. The small discrepancies are due to increased hybridisations and band inversion between p and d- states due to spin-orbit coupling in the `non-corrected' case.   

\begin{figure}
	\centering
	\includegraphics[width=0.95\columnwidth]{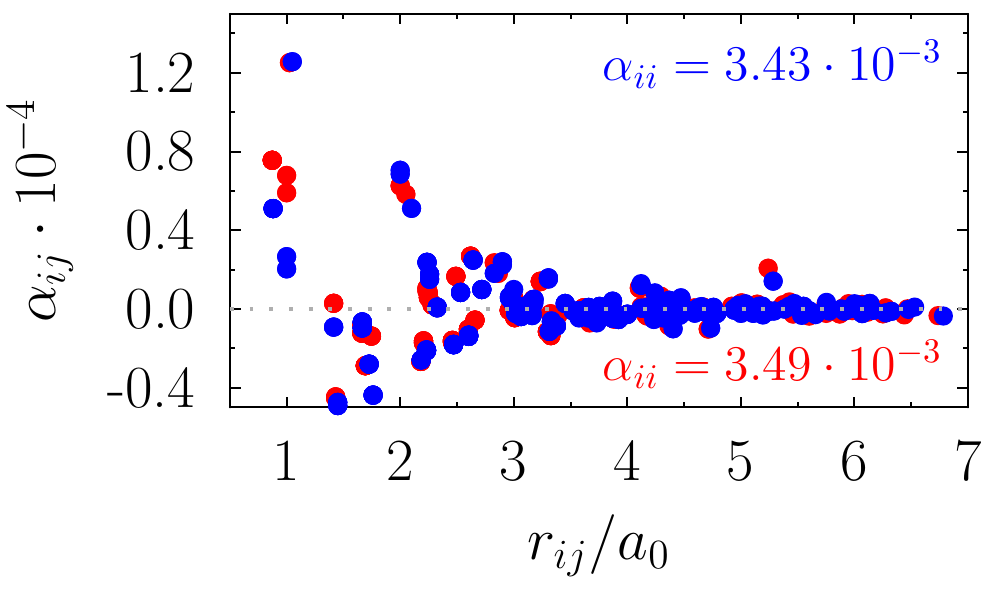}
	\caption{(Color online) Non-local Gilbert damping as a function of the normalized distance $\nicefrac{\vec{r}_{ij}}{a_0}$ for a tetragonal distorted bcc Fe crystal structure. Here, $\nicefrac{c}{a}=1.025$ (red circles) and $\nicefrac{c}{a}=1.05$ (blue circles) is considered. $\Gamma$ is put to $\unit[0.01]{eV}$. The zero value is indicated by dotted lines. }
	\label{fig:7}
\end{figure}

Finally, we address other ways than temperature (here represented by $\Gamma$), to manipulate the non-local damping. It is well established in literature already for Heisenberg exchange and the magneto crystalline anisotropy that compressive or tensial strain can be used to tune the magnetic phase stability and to design multiferroic materials. In an analogous way, also non-local damping depends on distortions in the crystal (see Fig.~\ref{fig:7}).  

Here, we applied non-volume conserved tetragonal strain along the $c$ axis. The local damping $\alpha_{ii}$ is marginal biased. Relative to the values of the undistorted case, a stronger effect is observed for the non-local part, in particular for the first few neighbours. Since we do a non-volume conserved distortion, the in-plane second NN component of the non-local damping is constant. The damping is in general decreasing with increasing distortion, however, a change in the sign of the damping can also occur (e.g. for the third NN). The rate of change in damping is not linear. In particular, the nearest-neighbour rate is about $\delta\alpha \approx 0.4\cdot10^{-5}$ for $2.5\%$ distortion, and $2.9\cdot10^{-5}$ for $5\%$ from the undistorted case. For the second nearest neighbour, the rate is even bigger ($3.0\cdot10^{-5}$ for $2.5\%$, $6.9\cdot10^{-5}$ for $5\%$). For neighbours larger than $r_{ij}=3a_0$, the change is less significant ($-0.6\cdot10^{-5}$ for $2.5\%$, $-0.7\cdot10^{-5}$ for $5\%$). The strongly strain dependent damping motivates even higher-order coupled damping contributions obtained from Taylor expanding the damping contribution around the equilibrium position $\alpha^0_{ij}$: $\alpha_{ij}=\alpha^0_{ij}+\nicefrac{\partial \alpha_{ij}}{\partial \vec{u}_k}\cdot \vec{u}_k + \hdots$. Note that this is in analogy to the magnetic exchange interaction \cite{Ma:2012hp} (exchange striction) and a natural name for it would be `dissipation striction'. This opens new ways to dissipatively couple spin and lattice reservoir in combined dynamics \cite{Ma:2012hp}, to the best of our knowledge not considered in todays ab-initio modelling of atomistic magnetisation dynamics.

\subsection{Atomistic magnetisation dynamics}
\begin{figure}
	\centering
	\includegraphics[width=0.95\columnwidth]{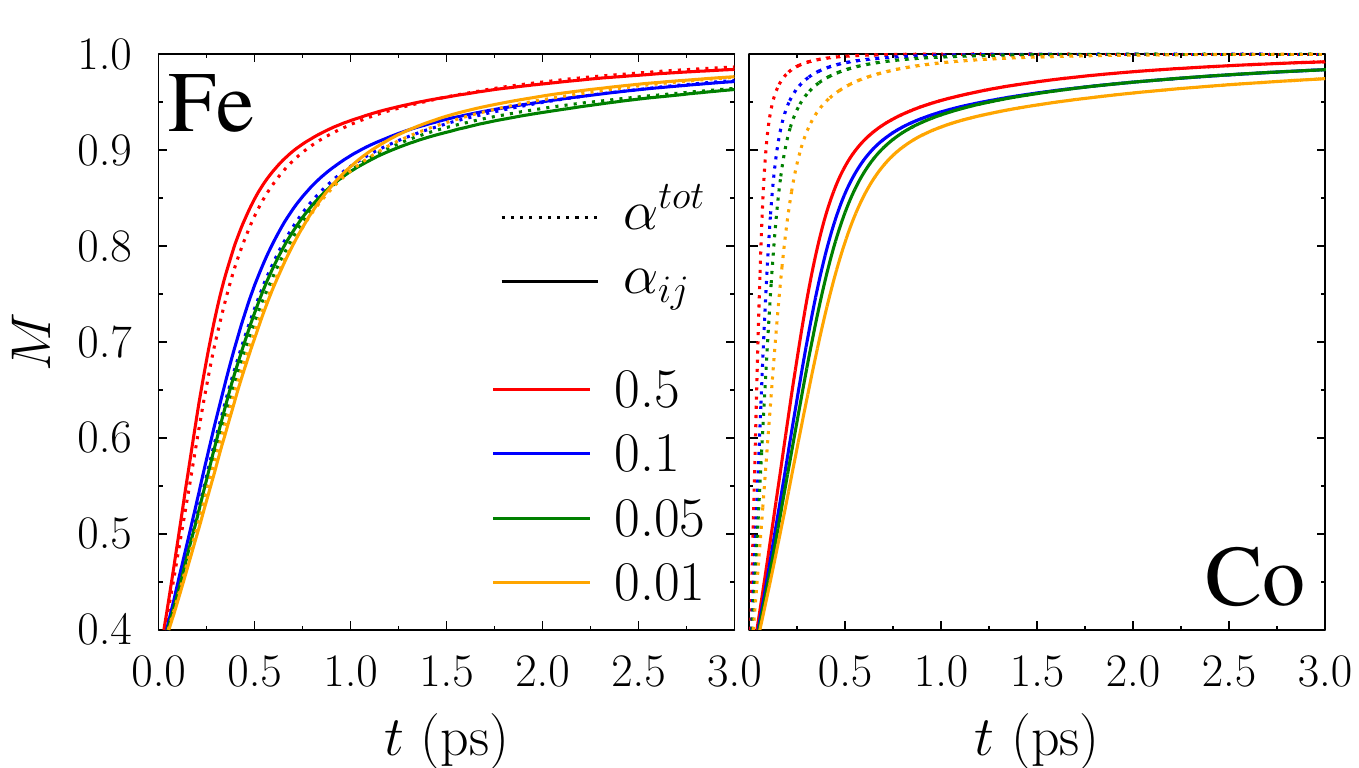}
	\caption{(Color online) Evolution of the average magnetic moment $M$ during remagnetization in bcc Fe (left panel) and fcc Co (right panel) for different damping strength according to the spectral width $\Gamma$ (different colors) and both, full non-local $\alpha_{ij}$ (solid line) and total, purely local $\alpha^{tot}$ (dashed line) Gilbert damping.}
	\label{fig:8}
\end{figure}

The question about the importance of non-local damping in atomistic magnetization dynamics (ASD) remains. For this purpose, we performed zero-temperature ASD for bcc Fe and fcc Co bulk and analysed changes in the average magnetization during relaxation from a totally random magnetic configuration, for which the total moment was zero (Fig. \ref{fig:8})

Related to the spectral width, the velocity for remagnetisation changes and is higher, the bigger the effective Gilbert damping is. For comparison, we performed also ASD simulations based on Eq.~\eqref{eq:1.1} with a scalar, purely local damping $\alpha^{tot}$ (dotted lines). For Fe, it turned out that accounting for the non-local damping causes a slight decrease in the remagnetization time, however, is overall not important for relaxation processes. This is understandable by comparing the particular damping values in Fig.~\ref{fig:4}, right panel, in which the non-local part appear negligible. On the other hand, for Co the effect on the relaxation process is much more significant, since the non-local Gilbert damping reduces the local contribution drastically (see Fig.~\ref{fig:4}, right panel). This `negative' non-local part ($i\neq j$) in $\alpha_{ij}$ decelerates the relaxation process and the relaxation time is drastically increased by a factor of $10$. Note that a `positive' non-local part will accelerate the relaxation, which is of high interest for ultrafast switching processes.

\section{Concluding remarks} 
\label{sec:concl}
In conclusion, we have evaluated the non-locality of the Gilbert damping parameter in both reciprocal and real space for elemental, itinerant magnets bcc Fe, fcc Co and fcc Ni. In particular in the reciprocal space, our results are in good agreement with values given in  the literature \cite{Gilmore:2009hm}. The here studied real space damping was considered on an atomistic level and it motivates to account for the full, non-local Gilbert damping in magnetization dynamic, e.g. at surfaces \cite{Bergqvist:2013iu} or for nano-structures \cite{Bottcher:2011dt}. We revealed that non-local damping can be negative, has a spatial anisotropy, quadratically scales with spin-orbit coupling, and decays in space as $r_{ij}^{-2}$. Detailed comparison between real and reciprocal states identified the importance of the corrections proposed by Edwards \cite{Edwards:2016be} and, consequently, overcome the limits of the Kambersk\'y formula showing an unphysical and experimental not proved divergent behaviour at low temperature. We further promote ways of manipulating non-local Gilbert damping, either by temperature, material´s doping or strain, and motivating `dissipation striction' terms, that opens a fundamental new root in the coupling between spin and lattice reservoirs. 

Our studies are the starting point for even further investigations: Although we mimic temperature by the spectral broadening $\Gamma$, a precise mapping of $\Gamma$ to spin and phonon temperature is still missing, according to Refs.~[\onlinecite{Thonig:2014kt,Ebert:2011gx}]. Even at zero temperature, we revealed a significant effect of the non-local Gilbert damping to the magnetization dynamics, but the influence of non-local damping to finite temperature analysis or even to low-dimensional structures has to be demonstrated.

\section{Acknowledgements}
\label{sec:acknow}
The authors thank Lars Bergqvist, Lars Nordstr\"om, Justin Shaw, and Jonas Fransson for fruitful discussions. O.E. acknowledges the support from  Swedish Research Council (VR), eSSENCE, and the KAW Foundation (Grants No. 2012.0031 and No. 2013.0020). 

\appendix
\section{Numerical details}
\label{sec:AppendixA}

We perform $\vec{k}$ integration with up to $1.25\cdot 10^6$ mesh points ($500\times 500\times 500$) in the first Brillouin zone for bulk. The energy integration is evaluated at the Fermi level only. For our principles studies, we performed a Slater-Koster parameterised \cite{Slater:1954hi} tight binding (TB) calculations~\cite{thonig-code} of the torque-torque correlation model as well as for the Green's function model. Here, the TB parameters have been obtained by fitting the electronic structures to those of a first-principles fully relativistic multiple scattering Korringa-Kohn-Rostoker (KKR) method using a genetic algorithm. The details of the fitting and the tight binding parameters are listed elsewhere \cite{Thonig:2014kt,Thonig:2017jf}. This puts our model on a firm, first-principles ground.

The tight binding Hamiltonian\cite{Schena:2010zg} $\mathcal{H}=\mathcal{H}_0+\mathcal{H}_{mag}+\mathcal{H}_{soc}$ contains on-site energies and hopping elements $\mathcal{H}_0$, the spin-orbit coupling $\mathcal{H}_{soc}=\zeta \vec{S}\cdot\vec{L}$  and the Zeeman term $\mathcal{H}_{mag}=\nicefrac{1}{2}\vec{B}\cdot\vec{\sigma}$. The Green's function is obtained by $\mathcal{G}=(\varepsilon+\cone \Gamma - \mathcal{H})^{-1}$, allows in principle to consider disorder in terms of spin and phonon as well as alloys \cite{Thonig:2014kt}. The bulk Greenian $\mathcal{G}_{ij}$ in real space between site $i$ and $j$ is obtained by Fourier transformation. Despite the fact that the tight binding approach is limited in accuracy, it produces good agreement with first principle band structure calculations for energies smaller than $ \varepsilon_F+\unit[5]{eV}$. 

\begin{figure}
	\centering
	\includegraphics[width=1.0\columnwidth]{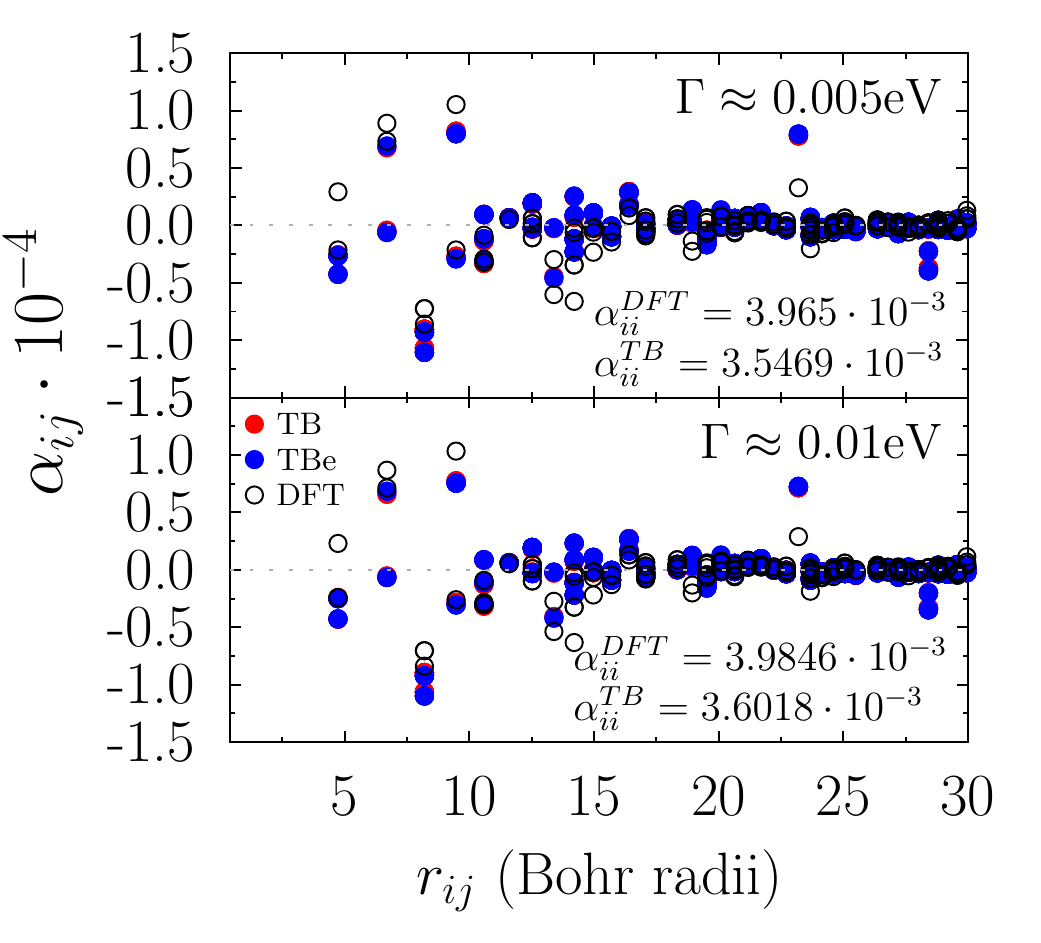}
	\caption{(Colour online) Comparison of non-local damping obtained from the Tight Binding method (TB) (red filled symbols), Tight Binding with Edwards correction (TBe) (blue filled symbols) and the linear muffin tin orbital method (DFT) (open symbols) for fcc Co. Two different spectral broadenings are chosen.}
	\label{fig:app:0}
\end{figure}

Equation~\eqref{eq:1.3} was also evaluated within the DFT and linear muffin-tin orbital method (LMTO) based code RSPt. The calculations were done for a k-point mesh of 128$^3$ k-points. We used three types of basis functions, characterised by different kinetic energies with $\kappa^2={0.1,-0.8,-1.7}$ Ry to describe $4s$, $4p$ and $3d$ states. The damping constants were calculated between the $3d$ orbitals, obtained using using muffin-tin head projection scheme \cite{Grechnev:2007en}. Both the first principles and tight binding implementation of the non-local Gilbert damping agree well (see Fig. \ref{fig:app:0}).

Note that due to numerical reasons, the values of $\Gamma$ used for the comparisons are slightly different in both electronic structure methods. Furthermore, in the LMTO method the orbitals are projected to d-orbitals only, which lead to small discrepancies in the damping.  

The atomistic magnetization dynamics is also performed within the Cahmd simulation package~\cite{thonig-code}. To reproduce bulk properties, periodic boundary conditions and a sufficiently large cluster $(10\times10\times10)$ are employed. The numerical time step is $\Delta t = \unit[0.1]{fs}$. The exchange coupling constants $J_{ij}$ are obtained from the Liechtenstein-Kastnelson-Antropov-Gubanovski (LKAG) formula implemented in the  first-principles fully relativistic multiple scattering Korringa-Kohn-Rostoker (KKR) method \cite{Zabloudil:2005dJ}. On the other hand, the magneto-crystalline anisotropy is used as a fixed parameter with $K=\unit[50]{\mu eV}$. 

\section{Spin-orbit coupling scaling in real and reciprocal space}
\label{sec:AppendixB}
\begin{figure}
	\centering
	\includegraphics[width=0.8\columnwidth]{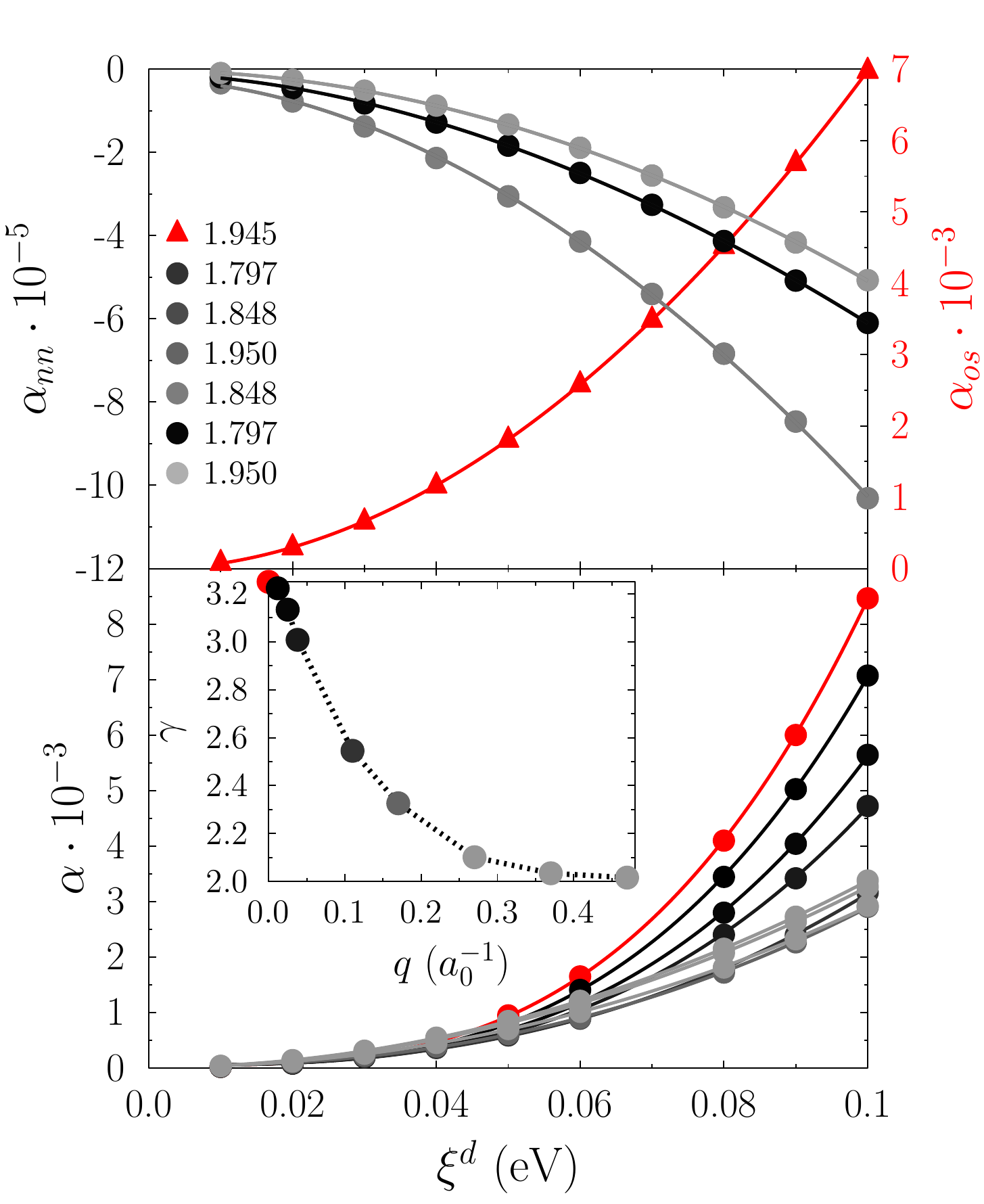}
	\caption{(Color online) Gilbert damping $\alpha$ as a function of the spin-orbit coupling for the d-states in fcc Co. Lower panel shows the Gilbert damping in reciprocal space for different $\vec{q}=\left|\vec{q}\right|$ values (different gray colours) along the $\Gamma \rightarrow X$ path. The upper panel exhibits the on-site $\alpha_{os}$ (red dotes and lines) and nearest-neighbour $\alpha_{nn}$ (gray dots and lines) damping. The solid line is the exponential fit of the data point. The inset shows the fitted exponents $\gamma$ with respect wave vector $q$. The colour of the dots is adjusted to the particular branch in the main figure. The spectral width is $\Gamma=\unit[0.005]{eV}$.}
	\label{fig:app:1}
\end{figure}

Kambersk\'y's formula is valid only for quadratic spin-orbit coupling scaling \cite{Costa:2015hp,Barati:2014gha}, which implies only scattering between states that preserve the spin. This mechanism was explicitly accounted by Edwards \cite{Edwards:2016be} by neglecting the spin-orbit coupling contribution in the `host' Green's function. It is predicted for the coherent mode ($\vec{q}=0$)\cite{Barati:2014gha} that this overcomes the unphysical and not experimentally verified divergent Gilbert damping for low temperature. Thus, the methodology requires to prove the functional dependency of the (non-local) Gilbert damping with respect to the spin-orbit coupling constant $\xi$ (Fig.~\ref{fig:app:1}). Since damping is a Fermi-surface effects, it is sufficient to consider only the spin-orbit coupling of the d-states. The real space Gilbert damping $\alpha_{ij}\propto \xi^\gamma$ scales for both on-site and nearest-neighbour sites with $\gamma\approx 2$. For the reciprocal space, however, the scaling is more complex and $\gamma$ depends on the magnon wave vector $\vec{q}$ (inset in Fig.~\ref{fig:app:1}). In the long-wavelength limit, the Kambersk\'y formula is valid, where for the ferromagnetic magnon mode with $\gamma\approx 3$ the Kambersk\'y formula is indefinite according to Edwards \cite{Edwards:2016be}.

\section{Intraband corrections}
\label{sec:AppendixC}
\begin{figure}
	\centering
	\includegraphics[width=1.0\columnwidth]{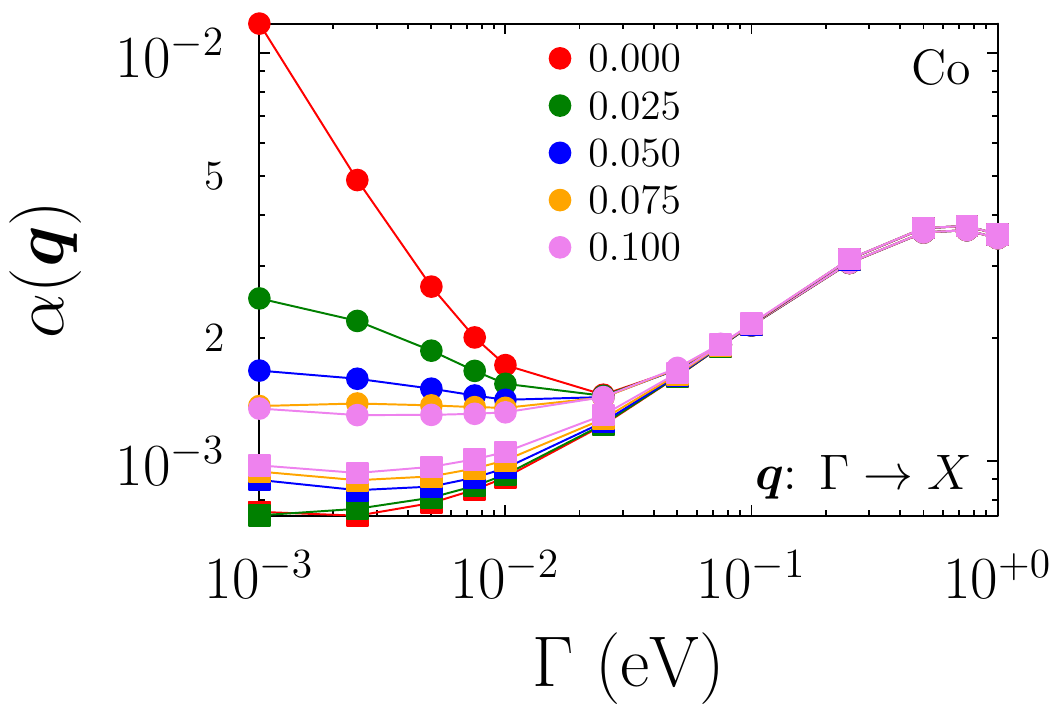}
	\caption{(Colour online) Comparison of reciprocal non-local damping with (squares) or without (circles) corrections proposed by Costa et al. \cite{Costa:2015hp} and Edwards \cite{Edwards:2016be} for Co and different spectral broadening $\Gamma$. Different colours represent different magnon propagation vectors $q$.}
	\label{fig:app:2}
\end{figure}

From the same reason as discussed in Section \ref{sec:AppendixB}, the role of the correction proposed by Edwards \cite{Edwards:2016be} for magnon propagations different than zero is unclear and need to be studied. Hence, we included the correction of Edward also to Eq.~\eqref{eq:1.2} (Fig.~\ref{fig:app:2}). The exclusion of the spin-orbit coupling (SOC) in the `host' clearly makes a major qualitative and quantitative change: Although the interband transitions are unaffected, interband transitions are mainly suppressed, as it was already discussed by Barati et al.\cite{Barati:2014gha}. However, the intraband contributions are not totally removed for small $\Gamma$. For very small scattering rates, the damping is constant. Opposite to the `non-corrected' Kambersk\'y formula, the increase of the magnon wave number $\vec{q}$ gives an increase in the non-local damping which is in agreement to the observation made by Yuan et al. \cite{Yuan:2014en}, but also with the analytical model proposed in Ref.~[\onlinecite{Umetsu:2012ga}] for small $\vec{q}$. This behaviour was observed for all itinerant magnets studied here.

\section{Comparison real and reciprocal Gilbert damping}
\label{sec:AppendixD}

\begin{figure}[h]
	\centering
	\includegraphics[width=0.95\columnwidth]{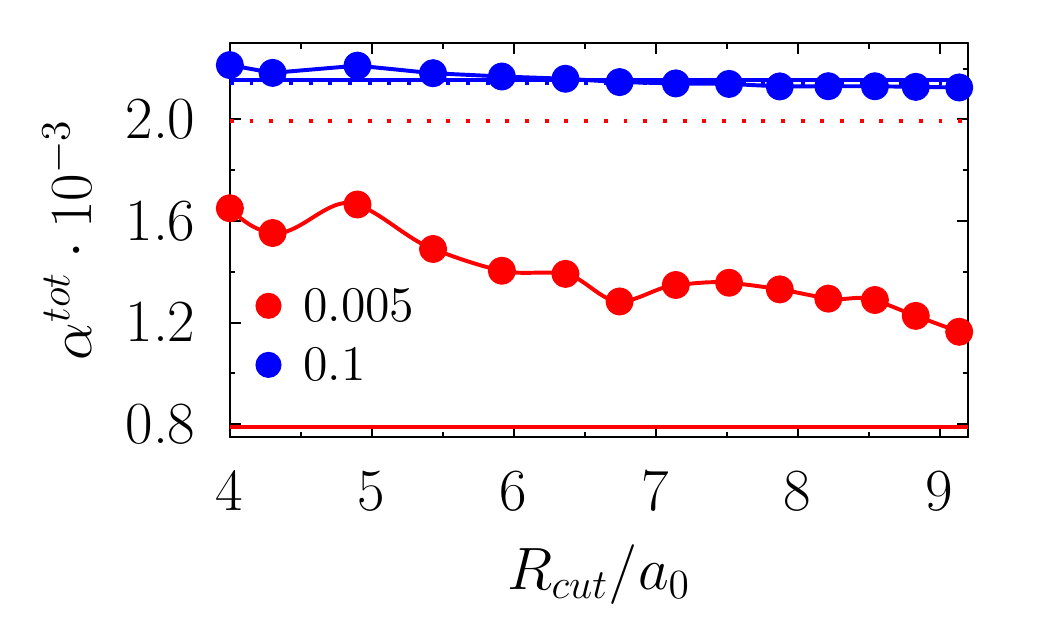}
	\caption{Total Gilbert damping $\alpha^{tot}$ for fcc Co as a function of summation cut-off radius for two spectral width $\Gamma$, one in intraband ($\Gamma=\unit[0.005]{eV}$, red dottes and lines) and one in the interband ($\Gamma=\unit[0.1]{eV}$, blue dottes and lines) region. The dotted and solid lines indicates the reciprocal value $\alpha(q=0)$ with and without SOC corrections, respectively.}
	\label{fig:app:3}
\end{figure}

The non-local damping scales like $r_{ij}^{-2}$ with the distance between the sites $i$ and $j$, and is, thus, very long range. In order to compare $\alpha^{tot}=\sum_{j\in R_{cut}}  \alpha_{ij}$ for arbitrary $i$ with $\alpha(\vec{q}=0)$, we have to specify the cut-off radius of the summation in real space (Fig.~\ref{fig:app:3}). The interband transitions ($\Gamma>\unit[0.05]{eV}$) are already converged for small cut-off radii $R_{cut}=3 a_0$. Intraband transitions, on the other hand, converge weakly with $R_{cut}$ to the reciprocal space value $\alpha(\vec{q}=0)$. Note that $\alpha(\vec{q}=0)$ is obtained from the corrected formalism. Even with $R_{cut}=9a_0$ which is proportional to $\approx 30000$ atoms, we have not obtain convergence.

\bibliographystyle{apsrev}

\begin{thebibliography}{61}
\expandafter\ifx\csname natexlab\endcsname\relax\def\natexlab#1{#1}\fi
\expandafter\ifx\csname bibnamefont\endcsname\relax
  \def\bibnamefont#1{#1}\fi
\expandafter\ifx\csname bibfnamefont\endcsname\relax
  \def\bibfnamefont#1{#1}\fi
\expandafter\ifx\csname citenamefont\endcsname\relax
  \def\citenamefont#1{#1}\fi
\expandafter\ifx\csname url\endcsname\relax
  \def\url#1{\texttt{#1}}\fi
\expandafter\ifx\csname urlprefix\endcsname\relax\def\urlprefix{URL }\fi
\providecommand{\bibinfo}[2]{#2}
\providecommand{\eprint}[2][]{\url{#2}}

\bibitem[{\citenamefont{Parkin et~al.}(2008)\citenamefont{Parkin, Hayashi, and
  Thomas}}]{Anonymous:mXuz0sqX}
\bibinfo{author}{\bibfnamefont{S.~S.~P.} \bibnamefont{Parkin}},
  \bibinfo{author}{\bibfnamefont{M.}~\bibnamefont{Hayashi}}, \bibnamefont{and}
  \bibinfo{author}{\bibfnamefont{L.}~\bibnamefont{Thomas}},
  \bibinfo{journal}{Science} \textbf{\bibinfo{volume}{320}},
  \bibinfo{pages}{190} (\bibinfo{year}{2008}),
  \urlprefix\url{http://www.sciencemag.org/cgi/doi/10.1126/science.1145799}.

\bibitem[{\citenamefont{Iwasaki et~al.}(2013)\citenamefont{Iwasaki, Mochizuki,
  and Nagaosa}}]{Iwasaki:2013hb}
\bibinfo{author}{\bibfnamefont{J.}~\bibnamefont{Iwasaki}},
  \bibinfo{author}{\bibfnamefont{M.}~\bibnamefont{Mochizuki}},
  \bibnamefont{and} \bibinfo{author}{\bibfnamefont{N.}~\bibnamefont{Nagaosa}},
  \bibinfo{journal}{Nature Nanotech} \textbf{\bibinfo{volume}{8}},
  \bibinfo{pages}{742} (\bibinfo{year}{2013}),
  \urlprefix\url{http://www.nature.com/doifinder/10.1038/nnano.2013.176}.

\bibitem[{\citenamefont{Fert et~al.}(2013)\citenamefont{Fert, Cros, and
  Sampaio}}]{Fert:2013fq}
\bibinfo{author}{\bibfnamefont{A.}~\bibnamefont{Fert}},
  \bibinfo{author}{\bibfnamefont{V.}~\bibnamefont{Cros}}, \bibnamefont{and}
  \bibinfo{author}{\bibfnamefont{J.}~\bibnamefont{Sampaio}},
  \bibinfo{journal}{Nature Nanotech} \textbf{\bibinfo{volume}{8}},
  \bibinfo{pages}{152} (\bibinfo{year}{2013}),
  \urlprefix\url{http://www.nature.com/doifinder/10.1038/nnano.2013.29}.

\bibitem[{\citenamefont{Behin-Aein et~al.}(2010)\citenamefont{Behin-Aein,
  Datta, Salahuddin, and Datta}}]{BehinAein:2010hj}
\bibinfo{author}{\bibfnamefont{B.}~\bibnamefont{Behin-Aein}},
  \bibinfo{author}{\bibfnamefont{D.}~\bibnamefont{Datta}},
  \bibinfo{author}{\bibfnamefont{S.}~\bibnamefont{Salahuddin}},
  \bibnamefont{and} \bibinfo{author}{\bibfnamefont{S.}~\bibnamefont{Datta}},
  \bibinfo{journal}{Nature Nanotech} \textbf{\bibinfo{volume}{5}},
  \bibinfo{pages}{266} (\bibinfo{year}{2010}),
  \urlprefix\url{http://www.nature.com/doifinder/10.1038/nnano.2010.31}.

\bibitem[{\citenamefont{Locatelli et~al.}()\citenamefont{Locatelli, Vincent,
  Mizrahi, Friedman, Vodenicarevic, Kim, Klein, Zhao, Grollier, and
  Querlioz}}]{Anonymous:jd}
\bibinfo{author}{\bibfnamefont{N.}~\bibnamefont{Locatelli}},
  \bibinfo{author}{\bibfnamefont{A.~F.} \bibnamefont{Vincent}},
  \bibinfo{author}{\bibfnamefont{A.}~\bibnamefont{Mizrahi}},
  \bibinfo{author}{\bibfnamefont{J.~S.} \bibnamefont{Friedman}},
  \bibinfo{author}{\bibfnamefont{D.}~\bibnamefont{Vodenicarevic}},
  \bibinfo{author}{\bibfnamefont{J.-V.} \bibnamefont{Kim}},
  \bibinfo{author}{\bibfnamefont{J.-O.} \bibnamefont{Klein}},
  \bibinfo{author}{\bibfnamefont{W.}~\bibnamefont{Zhao}},
  \bibinfo{author}{\bibfnamefont{J.}~\bibnamefont{Grollier}}, \bibnamefont{and}
  \bibinfo{author}{\bibfnamefont{D.}~\bibnamefont{Querlioz}}, in
  \emph{\bibinfo{booktitle}{DATE 2015}} (\bibinfo{publisher}{IEEE Conference
  Publications}, \bibinfo{address}{New Jersey}, ????), pp.
  \bibinfo{pages}{994--999}, ISBN \bibinfo{isbn}{9783981537048},
  \urlprefix\url{http://ieeexplore.ieee.org/xpl/articleDetails.jsp?arnumber=7092535}.

\bibitem[{\citenamefont{Koumpouras et~al.}(2017)\citenamefont{Koumpouras,
  Yudin, Adelmann, Bergman, Eriksson, and Pereiro}}]{Koumpouras:2017uc}
\bibinfo{author}{\bibfnamefont{K.}~\bibnamefont{Koumpouras}},
  \bibinfo{author}{\bibfnamefont{D.}~\bibnamefont{Yudin}},
  \bibinfo{author}{\bibfnamefont{C.}~\bibnamefont{Adelmann}},
  \bibinfo{author}{\bibfnamefont{A.}~\bibnamefont{Bergman}},
  \bibinfo{author}{\bibfnamefont{O.}~\bibnamefont{Eriksson}}, \bibnamefont{and}
  \bibinfo{author}{\bibfnamefont{M.}~\bibnamefont{Pereiro}}
  (\bibinfo{year}{2017}), \eprint{1702.00579},
  \urlprefix\url{https://arxiv.org/abs/1702.00579}.

\bibitem[{\citenamefont{Eriksson et~al.}(2016)\citenamefont{Eriksson, Bergman,
  Bergqvist, and Hellsvik}}]{Eriksson:2016uw}
\bibinfo{author}{\bibfnamefont{O.}~\bibnamefont{Eriksson}},
  \bibinfo{author}{\bibfnamefont{A.}~\bibnamefont{Bergman}},
  \bibinfo{author}{\bibfnamefont{L.}~\bibnamefont{Bergqvist}},
  \bibnamefont{and} \bibinfo{author}{\bibfnamefont{J.}~\bibnamefont{Hellsvik}},
  \emph{\bibinfo{title}{{Atomistic Spin Dynamics}}}, Foundations and
  Applications (\bibinfo{publisher}{Oxford University Press},
  \bibinfo{year}{2016}),
  \urlprefix\url{https://global.oup.com/academic/product/atomistic-spin-dynamics-9780198788669}.

\bibitem[{\citenamefont{Antropov et~al.}(1996)\citenamefont{Antropov,
  Katsnelson, Harmon, van Schilfgaarde, and Kusnezov}}]{Antropov:1996td}
\bibinfo{author}{\bibfnamefont{V.~P.} \bibnamefont{Antropov}},
  \bibinfo{author}{\bibfnamefont{M.~I.} \bibnamefont{Katsnelson}},
  \bibinfo{author}{\bibfnamefont{B.~N.} \bibnamefont{Harmon}},
  \bibinfo{author}{\bibfnamefont{M.}~\bibnamefont{van Schilfgaarde}},
  \bibnamefont{and} \bibinfo{author}{\bibfnamefont{D.}~\bibnamefont{Kusnezov}},
  \bibinfo{journal}{Phys. Rev. B} \textbf{\bibinfo{volume}{54}},
  \bibinfo{pages}{1019} (\bibinfo{year}{1996}),
  \urlprefix\url{http://link.aps.org/doi/10.1103/PhysRevB.54.1019}.

\bibitem[{\citenamefont{Kambersk{\'{y}}}(2007)}]{Kambersky:2007bp}
\bibinfo{author}{\bibfnamefont{V.}~\bibnamefont{Kambersk{\'{y}}}},
  \bibinfo{journal}{Phys. Rev. B} \textbf{\bibinfo{volume}{76}},
  \bibinfo{pages}{134416} (\bibinfo{year}{2007}),
  \urlprefix\url{http://link.aps.org/doi/10.1103/PhysRevB.76.134416}.

\bibitem[{\citenamefont{Kambersk{\'{y}}}(1984)}]{Kambersky:1984iz}
\bibinfo{author}{\bibfnamefont{V.}~\bibnamefont{Kambersk{\'{y}}}},
  \bibinfo{journal}{Czech J Phys} \textbf{\bibinfo{volume}{34}},
  \bibinfo{pages}{1111} (\bibinfo{year}{1984}),
  \urlprefix\url{https://link.springer.com/article/10.1007/BF01590106}.

\bibitem[{\citenamefont{Kambersk{\'{y}}}(1976)}]{Kambersky:1976gi}
\bibinfo{author}{\bibfnamefont{V.}~\bibnamefont{Kambersk{\'{y}}}},
  \bibinfo{journal}{Czech J Phys} \textbf{\bibinfo{volume}{26}},
  \bibinfo{pages}{1366} (\bibinfo{year}{1976}),
  \urlprefix\url{http://link.springer.com/10.1007/BF01587621}.

\bibitem[{\citenamefont{Gilmore et~al.}(2007)\citenamefont{Gilmore, Idzerda,
  and Stiles}}]{Gilmore:2007eva}
\bibinfo{author}{\bibfnamefont{K.}~\bibnamefont{Gilmore}},
  \bibinfo{author}{\bibfnamefont{Y.~U.} \bibnamefont{Idzerda}},
  \bibnamefont{and} \bibinfo{author}{\bibfnamefont{M.~D.}
  \bibnamefont{Stiles}}, \bibinfo{journal}{Phys. Rev. Lett.}
  \textbf{\bibinfo{volume}{99}}, \bibinfo{pages}{027204}
  (\bibinfo{year}{2007}),
  \urlprefix\url{http://link.aps.org/doi/10.1103/PhysRevLett.99.027204}.

\bibitem[{\citenamefont{Starikov et~al.}(2010)\citenamefont{Starikov, Kelly,
  Brataas, Tserkovnyak, and Bauer}}]{Starikov:2010cz}
\bibinfo{author}{\bibfnamefont{A.~A.} \bibnamefont{Starikov}},
  \bibinfo{author}{\bibfnamefont{P.~J.} \bibnamefont{Kelly}},
  \bibinfo{author}{\bibfnamefont{A.}~\bibnamefont{Brataas}},
  \bibinfo{author}{\bibfnamefont{Y.}~\bibnamefont{Tserkovnyak}},
  \bibnamefont{and} \bibinfo{author}{\bibfnamefont{G.~E.~W.}
  \bibnamefont{Bauer}}, \bibinfo{journal}{Phys. Rev. Lett.}
  \textbf{\bibinfo{volume}{105}}, \bibinfo{pages}{236601}
  (\bibinfo{year}{2010}),
  \urlprefix\url{http://link.aps.org/doi/10.1103/PhysRevLett.105.236601}.

\bibitem[{\citenamefont{Ebert et~al.}(2011)\citenamefont{Ebert, Mankovsky,
  K{\"o}dderitzsch, and Kelly}}]{Ebert:2011gx}
\bibinfo{author}{\bibfnamefont{H.}~\bibnamefont{Ebert}},
  \bibinfo{author}{\bibfnamefont{S.}~\bibnamefont{Mankovsky}},
  \bibinfo{author}{\bibfnamefont{D.}~\bibnamefont{K{\"o}dderitzsch}},
  \bibnamefont{and} \bibinfo{author}{\bibfnamefont{P.~J.} \bibnamefont{Kelly}},
  \bibinfo{journal}{Phys. Rev. Lett.} \textbf{\bibinfo{volume}{107}},
  \bibinfo{pages}{066603} (\bibinfo{year}{2011}),
  \urlprefix\url{http://link.aps.org/doi/10.1103/PhysRevLett.107.066603}.

\bibitem[{\citenamefont{Mankovsky et~al.}(2013)\citenamefont{Mankovsky,
  K{\"o}dderitzsch, Woltersdorf, and Ebert}}]{Mankovsky:2013ii}
\bibinfo{author}{\bibfnamefont{S.}~\bibnamefont{Mankovsky}},
  \bibinfo{author}{\bibfnamefont{D.}~\bibnamefont{K{\"o}dderitzsch}},
  \bibinfo{author}{\bibfnamefont{G.}~\bibnamefont{Woltersdorf}},
  \bibnamefont{and} \bibinfo{author}{\bibfnamefont{H.}~\bibnamefont{Ebert}},
  \bibinfo{journal}{Phys. Rev. B} \textbf{\bibinfo{volume}{87}},
  \bibinfo{pages}{014430} (\bibinfo{year}{2013}),
  \urlprefix\url{http://link.aps.org/doi/10.1103/PhysRevB.87.014430}.

\bibitem[{\citenamefont{Schoen et~al.}(2016)\citenamefont{Schoen, Thonig,
  Schneider, Silva, Nembach, Eriksson, Karis, and Shaw}}]{Schoen:2016gcc}
\bibinfo{author}{\bibfnamefont{M.~A.~W.} \bibnamefont{Schoen}},
  \bibinfo{author}{\bibfnamefont{D.}~\bibnamefont{Thonig}},
  \bibinfo{author}{\bibfnamefont{M.~L.} \bibnamefont{Schneider}},
  \bibinfo{author}{\bibfnamefont{T.~J.} \bibnamefont{Silva}},
  \bibinfo{author}{\bibfnamefont{H.~T.} \bibnamefont{Nembach}},
  \bibinfo{author}{\bibfnamefont{O.}~\bibnamefont{Eriksson}},
  \bibinfo{author}{\bibfnamefont{O.}~\bibnamefont{Karis}}, \bibnamefont{and}
  \bibinfo{author}{\bibfnamefont{J.~M.} \bibnamefont{Shaw}},
  \bibinfo{journal}{Nat Phys} \textbf{\bibinfo{volume}{12}},
  \bibinfo{pages}{839} (\bibinfo{year}{2016}),
  \urlprefix\url{http://www.nature.com/doifinder/10.1038/nphys3770}.

\bibitem[{\citenamefont{Chico et~al.}(2016)\citenamefont{Chico, Keshavarz,
  Kvashnin, Pereiro, Di~Marco, Etz, Eriksson, Bergman, and
  Bergqvist}}]{Chico:2016dy}
\bibinfo{author}{\bibfnamefont{J.}~\bibnamefont{Chico}},
  \bibinfo{author}{\bibfnamefont{S.}~\bibnamefont{Keshavarz}},
  \bibinfo{author}{\bibfnamefont{Y.}~\bibnamefont{Kvashnin}},
  \bibinfo{author}{\bibfnamefont{M.}~\bibnamefont{Pereiro}},
  \bibinfo{author}{\bibfnamefont{I.}~\bibnamefont{Di~Marco}},
  \bibinfo{author}{\bibfnamefont{C.}~\bibnamefont{Etz}},
  \bibinfo{author}{\bibfnamefont{O.}~\bibnamefont{Eriksson}},
  \bibinfo{author}{\bibfnamefont{A.}~\bibnamefont{Bergman}}, \bibnamefont{and}
  \bibinfo{author}{\bibfnamefont{L.}~\bibnamefont{Bergqvist}},
  \bibinfo{journal}{Phys. Rev. B} \textbf{\bibinfo{volume}{93}},
  \bibinfo{pages}{214439} (\bibinfo{year}{2016}),
  \urlprefix\url{http://link.aps.org/doi/10.1103/PhysRevB.93.214439}.

\bibitem[{\citenamefont{D{\"u}rrenfeld
  et~al.}(2015)\citenamefont{D{\"u}rrenfeld, Gerhard, Chico, Dumas, Ranjbar,
  Bergman, Bergqvist, Delin, Gould, Molenkamp et~al.}}]{Durrenfeld:2015vx}
\bibinfo{author}{\bibfnamefont{P.}~\bibnamefont{D{\"u}rrenfeld}},
  \bibinfo{author}{\bibfnamefont{F.}~\bibnamefont{Gerhard}},
  \bibinfo{author}{\bibfnamefont{J.}~\bibnamefont{Chico}},
  \bibinfo{author}{\bibfnamefont{R.~K.} \bibnamefont{Dumas}},
  \bibinfo{author}{\bibfnamefont{M.}~\bibnamefont{Ranjbar}},
  \bibinfo{author}{\bibfnamefont{A.}~\bibnamefont{Bergman}},
  \bibinfo{author}{\bibfnamefont{L.}~\bibnamefont{Bergqvist}},
  \bibinfo{author}{\bibfnamefont{A.}~\bibnamefont{Delin}},
  \bibinfo{author}{\bibfnamefont{C.}~\bibnamefont{Gould}},
  \bibinfo{author}{\bibfnamefont{L.~W.} \bibnamefont{Molenkamp}},
  \bibnamefont{et~al.} (\bibinfo{year}{2015}), \eprint{1510.01894},
  \urlprefix\url{http://arxiv.org/abs/1510.01894}.

\bibitem[{\citenamefont{Schoen et~al.}(2017{\natexlab{a}})\citenamefont{Schoen,
  Lucassen, Nembach, Silva, Koopmans, Back, and Shaw}}]{Schoen:2017kv}
\bibinfo{author}{\bibfnamefont{M.~A.~W.} \bibnamefont{Schoen}},
  \bibinfo{author}{\bibfnamefont{J.}~\bibnamefont{Lucassen}},
  \bibinfo{author}{\bibfnamefont{H.~T.} \bibnamefont{Nembach}},
  \bibinfo{author}{\bibfnamefont{T.~J.} \bibnamefont{Silva}},
  \bibinfo{author}{\bibfnamefont{B.}~\bibnamefont{Koopmans}},
  \bibinfo{author}{\bibfnamefont{C.~H.} \bibnamefont{Back}}, \bibnamefont{and}
  \bibinfo{author}{\bibfnamefont{J.~M.} \bibnamefont{Shaw}},
  \bibinfo{journal}{Phys. Rev. B} \textbf{\bibinfo{volume}{95}},
  \bibinfo{pages}{134410} (\bibinfo{year}{2017}{\natexlab{a}}),
  \urlprefix\url{http://link.aps.org/doi/10.1103/PhysRevB.95.134410}.

\bibitem[{\citenamefont{Schoen et~al.}(2017{\natexlab{b}})\citenamefont{Schoen,
  Lucassen, Nembach, Koopmans, Silva, Back, and Shaw}}]{Schoen:2017hj}
\bibinfo{author}{\bibfnamefont{M.~A.~W.} \bibnamefont{Schoen}},
  \bibinfo{author}{\bibfnamefont{J.}~\bibnamefont{Lucassen}},
  \bibinfo{author}{\bibfnamefont{H.~T.} \bibnamefont{Nembach}},
  \bibinfo{author}{\bibfnamefont{B.}~\bibnamefont{Koopmans}},
  \bibinfo{author}{\bibfnamefont{T.~J.} \bibnamefont{Silva}},
  \bibinfo{author}{\bibfnamefont{C.~H.} \bibnamefont{Back}}, \bibnamefont{and}
  \bibinfo{author}{\bibfnamefont{J.~M.} \bibnamefont{Shaw}},
  \bibinfo{journal}{Phys. Rev. B} \textbf{\bibinfo{volume}{95}},
  \bibinfo{pages}{134411} (\bibinfo{year}{2017}{\natexlab{b}}),
  \urlprefix\url{http://link.aps.org/doi/10.1103/PhysRevB.95.134411}.

\bibitem[{\citenamefont{Barati et~al.}(2014)\citenamefont{Barati, Cinal,
  Edwards, and Umerski}}]{Barati:2014gha}
\bibinfo{author}{\bibfnamefont{E.}~\bibnamefont{Barati}},
  \bibinfo{author}{\bibfnamefont{M.}~\bibnamefont{Cinal}},
  \bibinfo{author}{\bibfnamefont{D.~M.} \bibnamefont{Edwards}},
  \bibnamefont{and} \bibinfo{author}{\bibfnamefont{A.}~\bibnamefont{Umerski}},
  \bibinfo{journal}{Phys. Rev. B} \textbf{\bibinfo{volume}{90}},
  \bibinfo{pages}{014420} (\bibinfo{year}{2014}),
  \urlprefix\url{http://link.aps.org/doi/10.1103/PhysRevB.90.014420}.

\bibitem[{\citenamefont{Pan et~al.}(2016)\citenamefont{Pan, Chico, Hellsvik,
  Delin, Bergman, and Bergqvist}}]{Pan:2016gb}
\bibinfo{author}{\bibfnamefont{F.}~\bibnamefont{Pan}},
  \bibinfo{author}{\bibfnamefont{J.}~\bibnamefont{Chico}},
  \bibinfo{author}{\bibfnamefont{J.}~\bibnamefont{Hellsvik}},
  \bibinfo{author}{\bibfnamefont{A.}~\bibnamefont{Delin}},
  \bibinfo{author}{\bibfnamefont{A.}~\bibnamefont{Bergman}}, \bibnamefont{and}
  \bibinfo{author}{\bibfnamefont{L.}~\bibnamefont{Bergqvist}},
  \bibinfo{journal}{Phys. Rev. B} \textbf{\bibinfo{volume}{94}},
  \bibinfo{pages}{214410} (\bibinfo{year}{2016}),
  \urlprefix\url{https://link.aps.org/doi/10.1103/PhysRevB.94.214410}.

\bibitem[{\citenamefont{Thonig and Henk}(2014)}]{Thonig:2014kt}
\bibinfo{author}{\bibfnamefont{D.}~\bibnamefont{Thonig}} \bibnamefont{and}
  \bibinfo{author}{\bibfnamefont{J.}~\bibnamefont{Henk}}, \bibinfo{journal}{New
  J. Phys.} \textbf{\bibinfo{volume}{16}}, \bibinfo{pages}{013032}
  (\bibinfo{year}{2014}),
  \urlprefix\url{http://iopscience.iop.org/article/10.1088/1367-2630/16/1/013032}.

\bibitem[{\citenamefont{Ma and Seiler}(2017)}]{Ma:2017bd}
\bibinfo{author}{\bibfnamefont{Z.}~\bibnamefont{Ma}} \bibnamefont{and}
  \bibinfo{author}{\bibfnamefont{D.~G.} \bibnamefont{Seiler}},
  \emph{\bibinfo{title}{{Metrology and Diagnostic Techniques for
  Nanoelectronics}}} (\bibinfo{publisher}{Pan Stanford},
  \bibinfo{address}{Taylor {\&} Francis Group, 6000 Broken Sound Parkway NW,
  Suite 300, Boca Raton, FL 33487-2742}, \bibinfo{year}{2017}), ISBN
  \bibinfo{isbn}{9789814745086},
  \urlprefix\url{http://www.crcnetbase.com/doi/book/10.1201/9781315185385}.

\bibitem[{\citenamefont{Steiauf and F{\"a}hnle}(2005)}]{Steiauf:2005bv}
\bibinfo{author}{\bibfnamefont{D.}~\bibnamefont{Steiauf}} \bibnamefont{and}
  \bibinfo{author}{\bibfnamefont{M.}~\bibnamefont{F{\"a}hnle}},
  \bibinfo{journal}{Phys. Rev. B} \textbf{\bibinfo{volume}{72}},
  \bibinfo{pages}{064450} (\bibinfo{year}{2005}),
  \urlprefix\url{https://link.aps.org/doi/10.1103/PhysRevB.72.064450}.

\bibitem[{\citenamefont{Thonig et~al.}(2015)\citenamefont{Thonig, Henk, and
  Eriksson}}]{Thonig:2015ur}
\bibinfo{author}{\bibfnamefont{D.}~\bibnamefont{Thonig}},
  \bibinfo{author}{\bibfnamefont{J.}~\bibnamefont{Henk}}, \bibnamefont{and}
  \bibinfo{author}{\bibfnamefont{O.}~\bibnamefont{Eriksson}},
  \bibinfo{journal}{Phys. Rev. B} \textbf{\bibinfo{volume}{92}},
  \bibinfo{pages}{104403} (\bibinfo{year}{2015}),
  \urlprefix\url{http://link.aps.org/doi/10.1103/PhysRevB.92.104403}.

\bibitem[{\citenamefont{F{\"a}hnle and Steiauf}(2006)}]{Fahnle:2006cf}
\bibinfo{author}{\bibfnamefont{M.}~\bibnamefont{F{\"a}hnle}} \bibnamefont{and}
  \bibinfo{author}{\bibfnamefont{D.}~\bibnamefont{Steiauf}},
  \bibinfo{journal}{Phys. Rev. B} \textbf{\bibinfo{volume}{73}},
  \bibinfo{pages}{184427} (\bibinfo{year}{2006}),
  \urlprefix\url{https://link.aps.org/doi/10.1103/PhysRevB.73.184427}.

\bibitem[{\citenamefont{Minguzzi}(2015)}]{Minguzzi:2015jsd}
\bibinfo{author}{\bibfnamefont{E.}~\bibnamefont{Minguzzi}},
  \bibinfo{journal}{European Journal of Physics} \textbf{\bibinfo{volume}{36}},
  \bibinfo{pages}{035014} (\bibinfo{year}{2015}),
  \urlprefix\url{http://adsabs.harvard.edu/cgi-bin/nph-data_query?bibcode=2015EJPh...36c5014M&link_type=EJOURNAL}.

\bibitem[{\citenamefont{Yuan et~al.}(2014)\citenamefont{Yuan, Hals, Liu,
  Starikov, Brataas, and Kelly}}]{Yuan:2014en}
\bibinfo{author}{\bibfnamefont{Z.}~\bibnamefont{Yuan}},
  \bibinfo{author}{\bibfnamefont{K.~M.~D.} \bibnamefont{Hals}},
  \bibinfo{author}{\bibfnamefont{Y.}~\bibnamefont{Liu}},
  \bibinfo{author}{\bibfnamefont{A.~A.} \bibnamefont{Starikov}},
  \bibinfo{author}{\bibfnamefont{A.}~\bibnamefont{Brataas}}, \bibnamefont{and}
  \bibinfo{author}{\bibfnamefont{P.~J.} \bibnamefont{Kelly}},
  \bibinfo{journal}{Phys. Rev. Lett.} \textbf{\bibinfo{volume}{113}},
  \bibinfo{pages}{266603} (\bibinfo{year}{2014}),
  \urlprefix\url{http://link.aps.org/doi/10.1103/PhysRevLett.113.266603}.

\bibitem[{\citenamefont{Nembach et~al.}(2013)\citenamefont{Nembach, Shaw,
  Boone, and Silva}}]{Nembach:2013wh}
\bibinfo{author}{\bibfnamefont{H.}~\bibnamefont{Nembach}},
  \bibinfo{author}{\bibfnamefont{J.}~\bibnamefont{Shaw}},
  \bibinfo{author}{\bibfnamefont{C.}~\bibnamefont{Boone}}, \bibnamefont{and}
  \bibinfo{author}{\bibfnamefont{T.}~\bibnamefont{Silva}},
  \bibinfo{journal}{Phys. Rev. Lett.} \textbf{\bibinfo{volume}{110}},
  \bibinfo{pages}{117201} (\bibinfo{year}{2013}),
  \urlprefix\url{http://link.aps.org/doi/10.1103/PhysRevLett.110.117201}.

\bibitem[{\citenamefont{Bhattacharjee et~al.}(2012)\citenamefont{Bhattacharjee,
  Nordstr{\"o}m, and Fransson}}]{Bhattacharjee:2012if}
\bibinfo{author}{\bibfnamefont{S.}~\bibnamefont{Bhattacharjee}},
  \bibinfo{author}{\bibfnamefont{L.}~\bibnamefont{Nordstr{\"o}m}},
  \bibnamefont{and} \bibinfo{author}{\bibfnamefont{J.}~\bibnamefont{Fransson}},
  \bibinfo{journal}{Phys. Rev. Lett.} \textbf{\bibinfo{volume}{108}},
  \bibinfo{pages}{057204} (\bibinfo{year}{2012}),
  \urlprefix\url{http://link.aps.org/doi/10.1103/PhysRevLett.108.057204}.

\bibitem[{\citenamefont{Gilmore and Stiles}(2009)}]{Gilmore:2009hm}
\bibinfo{author}{\bibfnamefont{K.}~\bibnamefont{Gilmore}} \bibnamefont{and}
  \bibinfo{author}{\bibfnamefont{M.~D.} \bibnamefont{Stiles}},
  \bibinfo{journal}{Phys. Rev. B} \textbf{\bibinfo{volume}{79}},
  \bibinfo{pages}{132407} (\bibinfo{year}{2009}),
  \urlprefix\url{http://link.aps.org/doi/10.1103/PhysRevB.79.132407}.

\bibitem[{\citenamefont{Hals et~al.}(2009)\citenamefont{Hals, Nguyen, and
  Brataas}}]{Hals:2009gy}
\bibinfo{author}{\bibfnamefont{K.~M.~D.} \bibnamefont{Hals}},
  \bibinfo{author}{\bibfnamefont{A.~K.} \bibnamefont{Nguyen}},
  \bibnamefont{and} \bibinfo{author}{\bibfnamefont{A.}~\bibnamefont{Brataas}},
  \bibinfo{journal}{Phys. Rev. Lett.} \textbf{\bibinfo{volume}{102}},
  \bibinfo{pages}{256601} (\bibinfo{year}{2009}),
  \urlprefix\url{https://link.aps.org/doi/10.1103/PhysRevLett.102.256601}.

\bibitem[{\citenamefont{Brataas et~al.}(2011)\citenamefont{Brataas,
  Tserkovnyak, and Bauer}}]{Brataas:2011caa}
\bibinfo{author}{\bibfnamefont{A.}~\bibnamefont{Brataas}},
  \bibinfo{author}{\bibfnamefont{Y.}~\bibnamefont{Tserkovnyak}},
  \bibnamefont{and} \bibinfo{author}{\bibfnamefont{G.~E.~W.}
  \bibnamefont{Bauer}}, \bibinfo{journal}{Phys. Rev. B}
  \textbf{\bibinfo{volume}{84}}, \bibinfo{pages}{054416}
  (\bibinfo{year}{2011}),
  \urlprefix\url{https://link.aps.org/doi/10.1103/PhysRevB.84.054416}.

\bibitem[{\citenamefont{Chimata et~al.}(2017)\citenamefont{Chimata,
  Delczeg-Czirjak, Szilva, Cardias, Kvashnin, Pereiro, Mankovsky, Ebert,
  Thonig, Sanyal et~al.}}]{Chimata:2017iu}
\bibinfo{author}{\bibfnamefont{R.}~\bibnamefont{Chimata}},
  \bibinfo{author}{\bibfnamefont{E.~K.} \bibnamefont{Delczeg-Czirjak}},
  \bibinfo{author}{\bibfnamefont{A.}~\bibnamefont{Szilva}},
  \bibinfo{author}{\bibfnamefont{R.}~\bibnamefont{Cardias}},
  \bibinfo{author}{\bibfnamefont{Y.~O.} \bibnamefont{Kvashnin}},
  \bibinfo{author}{\bibfnamefont{M.}~\bibnamefont{Pereiro}},
  \bibinfo{author}{\bibfnamefont{S.}~\bibnamefont{Mankovsky}},
  \bibinfo{author}{\bibfnamefont{H.}~\bibnamefont{Ebert}},
  \bibinfo{author}{\bibfnamefont{D.}~\bibnamefont{Thonig}},
  \bibinfo{author}{\bibfnamefont{B.}~\bibnamefont{Sanyal}},
  \bibnamefont{et~al.}, \bibinfo{journal}{Phys. Rev. B}
  \textbf{\bibinfo{volume}{95}}, \bibinfo{pages}{214417}
  (\bibinfo{year}{2017}),
  \urlprefix\url{http://link.aps.org/doi/10.1103/PhysRevB.95.214417}.

\bibitem[{\citenamefont{Gilmore}(2008)}]{Gilmore2008:tw}
\bibinfo{author}{\bibfnamefont{K.}~\bibnamefont{Gilmore}}, Ph.D. thesis,
  \bibinfo{school}{scholarworks.montana.edu} (\bibinfo{year}{2008}),
  \urlprefix\url{http://scholarworks.montana.edu/xmlui/bitstream/handle/1/1336/GilmoreK1207.pdf?sequence=1}.

\bibitem[{\citenamefont{Gilmore et~al.}(2011)\citenamefont{Gilmore, Garate,
  MacDonald, and Stiles}}]{Gilmore:2011be}
\bibinfo{author}{\bibfnamefont{K.}~\bibnamefont{Gilmore}},
  \bibinfo{author}{\bibfnamefont{I.}~\bibnamefont{Garate}},
  \bibinfo{author}{\bibfnamefont{A.~H.} \bibnamefont{MacDonald}},
  \bibnamefont{and} \bibinfo{author}{\bibfnamefont{M.~D.}
  \bibnamefont{Stiles}}, \bibinfo{journal}{Phys. Rev. B}
  \textbf{\bibinfo{volume}{84}}, \bibinfo{pages}{224412}
  (\bibinfo{year}{2011}),
  \urlprefix\url{http://link.aps.org/doi/10.1103/PhysRevB.84.224412}.

\bibitem[{\citenamefont{Sayad et~al.}(2016)\citenamefont{Sayad, Rausch, and
  Potthoff}}]{Sayad:2016dw}
\bibinfo{author}{\bibfnamefont{M.}~\bibnamefont{Sayad}},
  \bibinfo{author}{\bibfnamefont{R.}~\bibnamefont{Rausch}}, \bibnamefont{and}
  \bibinfo{author}{\bibfnamefont{M.}~\bibnamefont{Potthoff}},
  \bibinfo{journal}{Phys. Rev. Lett.} \textbf{\bibinfo{volume}{117}},
  \bibinfo{pages}{127201} (\bibinfo{year}{2016}),
  \urlprefix\url{http://link.aps.org/doi/10.1103/PhysRevLett.117.127201}.

\bibitem[{\citenamefont{Zabloudil et~al.}(2006)\citenamefont{Zabloudil,
  Szunyogh, Hammerling, and Weinberger}}]{Zabloudil:2005dJ}
\bibinfo{author}{\bibfnamefont{J.}~\bibnamefont{Zabloudil}},
  \bibinfo{author}{\bibfnamefont{L.}~\bibnamefont{Szunyogh}},
  \bibinfo{author}{\bibfnamefont{R.}~\bibnamefont{Hammerling}},
  \bibnamefont{and}
  \bibinfo{author}{\bibfnamefont{P.}~\bibnamefont{Weinberger}},
  \emph{\bibinfo{title}{{Electron Scattering in Solid Matter}}}, A Theoretical
  and Computational Treatise (\bibinfo{year}{2006}),
  \urlprefix\url{http://bookzz.org/md5/190C5DE184B30E2B6898DE499DFB7D78}.

\bibitem[{\citenamefont{Slater and Koster}(1954)}]{Slater:1954hi}
\bibinfo{author}{\bibfnamefont{J.~C.} \bibnamefont{Slater}} \bibnamefont{and}
  \bibinfo{author}{\bibfnamefont{G.~F.} \bibnamefont{Koster}},
  \bibinfo{journal}{Phys. Rev.} \textbf{\bibinfo{volume}{94}},
  \bibinfo{pages}{1498} (\bibinfo{year}{1954}),
  \urlprefix\url{http://link.aps.org/doi/10.1103/PhysRev.94.1498}.

\bibitem[{\citenamefont{Wills and Cooper}(1987)}]{Wills:1987kc}
\bibinfo{author}{\bibfnamefont{J.~M.} \bibnamefont{Wills}} \bibnamefont{and}
  \bibinfo{author}{\bibfnamefont{B.~R.} \bibnamefont{Cooper}},
  \bibinfo{journal}{Phys. Rev. B} \textbf{\bibinfo{volume}{36}},
  \bibinfo{pages}{3809} (\bibinfo{year}{1987}),
  \urlprefix\url{https://link.aps.org/doi/10.1103/PhysRevB.36.3809}.

\bibitem[{\citenamefont{Dreyss{\'e}}(2000)}]{Dreysse:2000uv}
\bibinfo{author}{\bibfnamefont{H.}~\bibnamefont{Dreyss{\'e}}},
  \emph{\bibinfo{title}{{Electronic Structure and Physical Properties of Solids
  }}} (\bibinfo{publisher}{Springer}, \bibinfo{year}{2000}),
  \urlprefix\url{http://bookzz.org/md5/66B5CD9A4859B7F039CB877263C4C9DD}.

\bibitem[{\citenamefont{Vittoria et~al.}(2010)\citenamefont{Vittoria, Yoon, and
  Widom}}]{Vittoria:2010ft}
\bibinfo{author}{\bibfnamefont{C.}~\bibnamefont{Vittoria}},
  \bibinfo{author}{\bibfnamefont{S.~D.} \bibnamefont{Yoon}}, \bibnamefont{and}
  \bibinfo{author}{\bibfnamefont{A.}~\bibnamefont{Widom}},
  \bibinfo{journal}{Phys. Rev. B} \textbf{\bibinfo{volume}{81}},
  \bibinfo{pages}{014412} (\bibinfo{year}{2010}),
  \urlprefix\url{https://link.aps.org/doi/10.1103/PhysRevB.81.014412}.

\bibitem[{\citenamefont{B{\"o}ttcher et~al.}(2012)\citenamefont{B{\"o}ttcher,
  Ernst, and Henk}}]{Bottcher:2012hz}
\bibinfo{author}{\bibfnamefont{D.}~\bibnamefont{B{\"o}ttcher}},
  \bibinfo{author}{\bibfnamefont{A.}~\bibnamefont{Ernst}}, \bibnamefont{and}
  \bibinfo{author}{\bibfnamefont{J.}~\bibnamefont{Henk}},
  \bibinfo{journal}{Journal of Magnetism and Magnetic Materials}
  \textbf{\bibinfo{volume}{324}}, \bibinfo{pages}{610} (\bibinfo{year}{2012}),
  \urlprefix\url{http://linkinghub.elsevier.com/retrieve/pii/S0304885311006299}.

\bibitem[{\citenamefont{Szilva et~al.}(2013)\citenamefont{Szilva, Costa,
  Bergman, Szunyogh, Nordstr{\"o}m, and Eriksson}}]{Szilva:2013fo}
\bibinfo{author}{\bibfnamefont{A.}~\bibnamefont{Szilva}},
  \bibinfo{author}{\bibfnamefont{M.}~\bibnamefont{Costa}},
  \bibinfo{author}{\bibfnamefont{A.}~\bibnamefont{Bergman}},
  \bibinfo{author}{\bibfnamefont{L.}~\bibnamefont{Szunyogh}},
  \bibinfo{author}{\bibfnamefont{L.}~\bibnamefont{Nordstr{\"o}m}},
  \bibnamefont{and} \bibinfo{author}{\bibfnamefont{O.}~\bibnamefont{Eriksson}},
  \bibinfo{journal}{Phys. Rev. Lett.} \textbf{\bibinfo{volume}{111}},
  \bibinfo{pages}{127204} (\bibinfo{year}{2013}),
  \urlprefix\url{http://link.aps.org/doi/10.1103/PhysRevLett.111.127204}.

\bibitem[{\citenamefont{Garate et~al.}(2009)\citenamefont{Garate, Gilmore,
  Stiles, and MacDonald}}]{Garate:2009ija}
\bibinfo{author}{\bibfnamefont{I.}~\bibnamefont{Garate}},
  \bibinfo{author}{\bibfnamefont{K.}~\bibnamefont{Gilmore}},
  \bibinfo{author}{\bibfnamefont{M.~D.} \bibnamefont{Stiles}},
  \bibnamefont{and} \bibinfo{author}{\bibfnamefont{A.~H.}
  \bibnamefont{MacDonald}}, \bibinfo{journal}{Phys. Rev. B}
  \textbf{\bibinfo{volume}{79}}, \bibinfo{pages}{104416}
  (\bibinfo{year}{2009}),
  \urlprefix\url{http://link.aps.org/doi/10.1103/PhysRevB.79.104416}.

\bibitem[{\citenamefont{Costa and Muniz}(2015)}]{Costa:2015hp}
\bibinfo{author}{\bibfnamefont{A.~T.} \bibnamefont{Costa}} \bibnamefont{and}
  \bibinfo{author}{\bibfnamefont{R.~B.} \bibnamefont{Muniz}},
  \bibinfo{journal}{Phys. Rev. B} \textbf{\bibinfo{volume}{92}},
  \bibinfo{pages}{014419} (\bibinfo{year}{2015}),
  \urlprefix\url{http://link.aps.org/doi/10.1103/PhysRevB.92.014419}.

\bibitem[{\citenamefont{Edwards}(2016)}]{Edwards:2016be}
\bibinfo{author}{\bibfnamefont{D.~M.} \bibnamefont{Edwards}},
  \bibinfo{journal}{J. Phys.: Condens. Matter} \textbf{\bibinfo{volume}{28}},
  \bibinfo{pages}{086004} (\bibinfo{year}{2016}),
  \urlprefix\url{http://iopscience.iop.org/article/10.1088/0953-8984/28/8/086004}.

\bibitem[{\citenamefont{Oogane et~al.}(2006)\citenamefont{Oogane, Wakitani,
  Yakata, Yilgin, Ando, Sakuma, and Miyazaki}}]{Oogane:2006bz}
\bibinfo{author}{\bibfnamefont{M.}~\bibnamefont{Oogane}},
  \bibinfo{author}{\bibfnamefont{T.}~\bibnamefont{Wakitani}},
  \bibinfo{author}{\bibfnamefont{S.}~\bibnamefont{Yakata}},
  \bibinfo{author}{\bibfnamefont{R.}~\bibnamefont{Yilgin}},
  \bibinfo{author}{\bibfnamefont{Y.}~\bibnamefont{Ando}},
  \bibinfo{author}{\bibfnamefont{A.}~\bibnamefont{Sakuma}}, \bibnamefont{and}
  \bibinfo{author}{\bibfnamefont{T.}~\bibnamefont{Miyazaki}},
  \bibinfo{journal}{Jpn. J. Appl. Phys.} \textbf{\bibinfo{volume}{45}},
  \bibinfo{pages}{3889} (\bibinfo{year}{2006}),
  \urlprefix\url{http://iopscience.iop.org/article/10.1143/JJAP.45.3889}.

\bibitem[{\citenamefont{Bhagat and Lubitz}(1974)}]{Bhagat:1974iu}
\bibinfo{author}{\bibfnamefont{S.~M.} \bibnamefont{Bhagat}} \bibnamefont{and}
  \bibinfo{author}{\bibfnamefont{P.}~\bibnamefont{Lubitz}},
  \bibinfo{journal}{Phys. Rev. B} \textbf{\bibinfo{volume}{10}},
  \bibinfo{pages}{179} (\bibinfo{year}{1974}),
  \urlprefix\url{https://link.aps.org/doi/10.1103/PhysRevB.10.179}.

\bibitem[{\citenamefont{Gilbert}(2004)}]{Gilbert:2004jg}
\bibinfo{author}{\bibfnamefont{T.~L.} \bibnamefont{Gilbert}},
  \bibinfo{journal}{IEEE Trans. Magn.} \textbf{\bibinfo{volume}{40}},
  \bibinfo{pages}{3443} (\bibinfo{year}{2004}),
  \urlprefix\url{http://adsabs.harvard.edu/cgi-bin/nph-data_query?bibcode=2004ITM....40.3443G&link_type=EJOURNAL}.

\bibitem[{\citenamefont{Umetsu et~al.}(2012)\citenamefont{Umetsu, Miura, and
  Sakuma}}]{Umetsu:2012ga}
\bibinfo{author}{\bibfnamefont{N.}~\bibnamefont{Umetsu}},
  \bibinfo{author}{\bibfnamefont{D.}~\bibnamefont{Miura}}, \bibnamefont{and}
  \bibinfo{author}{\bibfnamefont{A.}~\bibnamefont{Sakuma}},
  \bibinfo{journal}{J. Phys. Soc. Jpn.} \textbf{\bibinfo{volume}{81}},
  \bibinfo{pages}{114716} (\bibinfo{year}{2012}),
  \urlprefix\url{http://journals.jps.jp/doi/10.1143/JPSJ.81.114716}.

\bibitem[{\citenamefont{Pajda et~al.}(2001)\citenamefont{Pajda,
  Kudrnovsk{\'{y}}, Turek, Drchal, and Bruno}}]{Pajda:2001ix}
\bibinfo{author}{\bibfnamefont{M.}~\bibnamefont{Pajda}},
  \bibinfo{author}{\bibfnamefont{J.}~\bibnamefont{Kudrnovsk{\'{y}}}},
  \bibinfo{author}{\bibfnamefont{I.}~\bibnamefont{Turek}},
  \bibinfo{author}{\bibfnamefont{V.}~\bibnamefont{Drchal}}, \bibnamefont{and}
  \bibinfo{author}{\bibfnamefont{P.}~\bibnamefont{Bruno}},
  \bibinfo{journal}{Phys. Rev. B} \textbf{\bibinfo{volume}{64}},
  \bibinfo{pages}{174402} (\bibinfo{year}{2001}),
  \urlprefix\url{https://link.aps.org/doi/10.1103/PhysRevB.64.174402}.

\bibitem[{\citenamefont{Liechtenstein et~al.}(1987)\citenamefont{Liechtenstein,
  Katsnelson, Antropov, and Gubanov}}]{Liechtenstein:1987br}
\bibinfo{author}{\bibfnamefont{A.~I.} \bibnamefont{Liechtenstein}},
  \bibinfo{author}{\bibfnamefont{M.~I.} \bibnamefont{Katsnelson}},
  \bibinfo{author}{\bibfnamefont{V.~P.} \bibnamefont{Antropov}},
  \bibnamefont{and} \bibinfo{author}{\bibfnamefont{V.~A.}
  \bibnamefont{Gubanov}}, \bibinfo{journal}{Journal of Magnetism and Magnetic
  Materials} \textbf{\bibinfo{volume}{67}}, \bibinfo{pages}{65}
  (\bibinfo{year}{1987}),
  \urlprefix\url{http://linkinghub.elsevier.com/retrieve/pii/0304885387907219}.

\bibitem[{\citenamefont{Ma et~al.}(2012)\citenamefont{Ma, Dudarev, and
  Woo}}]{Ma:2012hp}
\bibinfo{author}{\bibfnamefont{P.-W.} \bibnamefont{Ma}},
  \bibinfo{author}{\bibfnamefont{S.~L.} \bibnamefont{Dudarev}},
  \bibnamefont{and} \bibinfo{author}{\bibfnamefont{C.~H.} \bibnamefont{Woo}},
  \bibinfo{journal}{Phys. Rev. B} \textbf{\bibinfo{volume}{85}},
  \bibinfo{pages}{184301} (\bibinfo{year}{2012}),
  \urlprefix\url{http://link.aps.org/doi/10.1103/PhysRevB.85.184301}.

\bibitem[{\citenamefont{Bergqvist et~al.}(2013)\citenamefont{Bergqvist, Taroni,
  Bergman, Etz, and Eriksson}}]{Bergqvist:2013iu}
\bibinfo{author}{\bibfnamefont{L.}~\bibnamefont{Bergqvist}},
  \bibinfo{author}{\bibfnamefont{A.}~\bibnamefont{Taroni}},
  \bibinfo{author}{\bibfnamefont{A.}~\bibnamefont{Bergman}},
  \bibinfo{author}{\bibfnamefont{C.}~\bibnamefont{Etz}}, \bibnamefont{and}
  \bibinfo{author}{\bibfnamefont{O.}~\bibnamefont{Eriksson}},
  \bibinfo{journal}{Phys. Rev. B} \textbf{\bibinfo{volume}{87}},
  \bibinfo{pages}{144401} (\bibinfo{year}{2013}),
  \urlprefix\url{http://link.aps.org/doi/10.1103/PhysRevB.87.144401}.

\bibitem[{\citenamefont{B{\"o}ttcher et~al.}(2011)\citenamefont{B{\"o}ttcher,
  Ernst, and Henk}}]{Bottcher:2011dt}
\bibinfo{author}{\bibfnamefont{D.}~\bibnamefont{B{\"o}ttcher}},
  \bibinfo{author}{\bibfnamefont{A.}~\bibnamefont{Ernst}}, \bibnamefont{and}
  \bibinfo{author}{\bibfnamefont{J.}~\bibnamefont{Henk}}, \bibinfo{journal}{J.
  Phys.: Condens. Matter} \textbf{\bibinfo{volume}{23}},
  \bibinfo{pages}{296003} (\bibinfo{year}{2011}),
  \urlprefix\url{http://iopscience.iop.org/article/10.1088/0953-8984/23/29/296003}.

\bibitem[{\citenamefont{Thonig}(2013)}]{thonig-code}
\bibinfo{author}{\bibfnamefont{D.}~\bibnamefont{Thonig}},
  \emph{\bibinfo{title}{{CaHmd - A computer program package for atomistic
  magnetisation dynamics simulations.}}}, \bibinfo{organization}{Uppsala
  University}, \bibinfo{address}{Uppsala}, \bibinfo{edition}{2nd} ed.
  (\bibinfo{year}{2013}).

\bibitem[{\citenamefont{Thonig et~al.}(2017)\citenamefont{Thonig, Eriksson, and
  Pereiro}}]{Thonig:2017jf}
\bibinfo{author}{\bibfnamefont{D.}~\bibnamefont{Thonig}},
  \bibinfo{author}{\bibfnamefont{O.}~\bibnamefont{Eriksson}}, \bibnamefont{and}
  \bibinfo{author}{\bibfnamefont{M.}~\bibnamefont{Pereiro}},
  \bibinfo{journal}{Sci. Rep.} \textbf{\bibinfo{volume}{7}},
  \bibinfo{pages}{931} (\bibinfo{year}{2017}),
  \urlprefix\url{http://www.nature.com/articles/s41598-017-01081-z}.

\bibitem[{\citenamefont{Schena}(2010)}]{Schena:2010zg}
\bibinfo{author}{\bibfnamefont{T.}~\bibnamefont{Schena}}, Ph.D. thesis,
  \bibinfo{school}{fz-juelich.de} (\bibinfo{year}{2010}),
  \urlprefix\url{http://www.fz-juelich.de/SharedDocs/Downloads/PGI/PGI-1/EN/Schena_diploma_pdf.pdf?__blob=publicationFile}.

\bibitem[{\citenamefont{Grechnev et~al.}(2007)\citenamefont{Grechnev, Di~Marco,
  Katsnelson, Lichtenstein, Wills, and Eriksson}}]{Grechnev:2007en}
\bibinfo{author}{\bibfnamefont{A.}~\bibnamefont{Grechnev}},
  \bibinfo{author}{\bibfnamefont{I.}~\bibnamefont{Di~Marco}},
  \bibinfo{author}{\bibfnamefont{M.~I.} \bibnamefont{Katsnelson}},
  \bibinfo{author}{\bibfnamefont{A.~I.} \bibnamefont{Lichtenstein}},
  \bibinfo{author}{\bibfnamefont{J.}~\bibnamefont{Wills}}, \bibnamefont{and}
  \bibinfo{author}{\bibfnamefont{O.}~\bibnamefont{Eriksson}},
  \bibinfo{journal}{Phys. Rev. B} \textbf{\bibinfo{volume}{76}},
  \bibinfo{pages}{035107} (\bibinfo{year}{2007}),
  \urlprefix\url{https://link.aps.org/doi/10.1103/PhysRevB.76.035107}.

\end{thebibliography}

\end{document}